\documentclass[aps,prd,amsmath
,eqsecnum            
,preprintnumbers
,nofootinbib         
 ,showpacs          
 ,showkeys          
,twocolumn         
]{revtex4}
\usepackage[dvips]{graphicx}
\usepackage{amsmath}
\usepackage{amsfonts}
\usepackage{amssymb}
\usepackage{longtable}   

\allowdisplaybreaks[4]     

\newcommand{\df}{\ {\overset {\rm def} =}\ }
\newcommand{\dr}[2]{\frac {{\rm d} {#1}} {{\rm d} {#2}}}

\newcommand{\dril}[2]{{{\rm d} {#1}} / {{\rm d} {#2}}}

\begin{document}

\title{Mimicking acceleration in the constant-bang-time Lema\^{\i}tre -- Tolman
model: \\
Shell crossings, density distributions and light cones}

\author{Andrzej Krasi\'nski}
\affiliation{N. Copernicus Astronomical Centre, Polish Academy of Sciences, \\
Bartycka 18, 00 716 Warszawa, Poland} \email{akr@camk.edu.pl}

\date {}

\begin{abstract}
The Lema\^{\i}tre -- Tolman model with $\Lambda = 0$ and constant bang time that
imitates the luminosity distance -- redshift relation of the $\Lambda$CDM model
using the energy function $E$ alone contains shell crossings. In this paper, the
location in spacetime and the consequences of existence of the shell-crossing
set (SCS) are investigated. The SCS would come into view of the central observer
only at $t \approx 1064 T$ to the future from now, where $T$ is the present age
of the Universe, but would not leave any recognizable trace in her observations.
Light rays emitted near to the SCS are blueshifted at the initial points, but
the blueshift is finite, and is overcompensated by later-induced redshifts if
the observer is sufficiently far. The local blueshifts cause that $z$ along a
light ray is not a monotonic function of the comoving radial coordinate $r$. As
a consequence, the angular diameter distance $D_A$ and the luminosity distance
$D_L$ from the central observer fail to be functions of $z$; the relations
$D_A(z)$ and $D_L(z)$ are multiple-valued in a vicinity of the SCS. The
following quantities are calculated and displayed: (1) The distribution of mass
density on a few characteristic hypersurfaces of constant time; some of them
intersect the SCS. (2) The distribution of density along the past light cone of
the present central observer. (3) A few light cones intersecting the SCS at
characteristic instants. (4) The redshift profiles along several light cones.
(5) The extremum-redshift hypersurface. (6) The $D_A(z)$ and $D_L(z)$ relations.
(7) The last scattering time and its comparison with the $\Lambda$CDM last
scattering epoch.
\end{abstract}

\maketitle

\section{Motivation and overview}\label{motivation}

\setcounter{equation}{0}

In Ref. \cite{Kras2014} it was shown how the luminosity distance -- redshift
relation $D_L(z)$ of the $\Lambda$CDM model is duplicated in the Lema\^{\i}tre
\cite{Lema1933} -- Tolman \cite{Tolm1934} (L--T) model with $\Lambda = 0$,
constant bang-time function $t_B$ and the energy function $E$ mimicking
accelerated expansion on the past light cone of the present central observer
(PCPO). This model was first introduced in Ref. \cite{INNa2002}, and further
investigated in Ref. \cite{YKNa2008}. Numerical calculations in Ref.
\cite{Kras2014} revealed that this model necessarily contains shell crossings in
the region whose boundary intersects the PCPO at $z = z_{\rm sc} \approx$
$6.938$. This is far enough to avoid any problems with the observations of the
type Ia supernovae \cite{Ries1998,Perl1999,Jone2014}, the farthest of which are
at $z_{\rm far} = 1.914$ \cite{Jone2014}. The shell crossings can be removed
from the model by matching it to a background (Friedmann, for example), at the
radial coordinate $r = r_m$, where the redshift corresponding to $r_m$ at the
PCPO is smaller than $z_{\rm sc}$, but larger than $z_{\rm far}$
\cite{Kras2014}.

In the present paper, the position of the shell-crossing set (SCS) relative to
the PCPO in that L--T model, and the consequences of its existence, are
investigated. The SCS lies so far to the future of the present time that it has
no influence on any observations that the present central observer could have
carried out until now (see below). Consequently, even if not removed, the shell
crossing is invisible for her. However, the position of the SCS in spacetime is
a consequence of the values of parameters of this particular model. With a
different shape of the $E(r)$ function, perhaps even with different values of
the $\Omega$ parameters, the SCS might appear early enough to be visible to the
present observer. In such a model, the findings about the redshift profiles
presented here will become relevant for the present-day astrophysics.

Sections \ref{basform} and \ref{useform} provide the basic formulae for
reference, extracted from Ref. \cite{Kras2014}. In Sec. \ref{lightcone}, the
profile and properties of the PCPO are presented. The location of the SCS in
spacetime is determined in Sec. \ref{locateSCS}.

In Sec. \ref{densities}, the distributions of mass density on a few
characteristic hypersurfaces of constant time are displayed; some of them
intersect the SCS. The distribution at the present time is, in agreement with
common expectation \cite{CBKr2010,BCKr2011}, that of a void, but with a cusp
rather than a smooth minimum at the center.\footnote{In some papers
\cite{VFWa2006} such a cusp was (mis)named a ``weak singularity'', although in
truth there is no singularity there \cite{KHBC2010}.\label{fnote1}} In Sec.
\ref{densoncone}, the distribution of density along the PCPO is displayed.

In Sec. \ref{SCSinflu}, the earliest ray emitted at the SCS that will reach the
central observer is determined. The observer will receive it at approximately
1064 $T$ in the future, where $T = 13.819 \times 10^9$ y is the observationally
determined present age of the Universe \cite{Plan2013}. Two other light cones
that intersect the SCS are calculated and displayed. As expected, in the
comoving coordinates they are horizontal at the intersection points.

In Sec. \ref{redprof}, redshift profiles along several past light cones reaching
the central observer are displayed. It is noted that on rays passing near the
SCS the function $z(r)$ has a local maximum at a certain $r$ and becomes
decreasing for larger $r$. This means that rays emitted near or at the SCS are
blueshifted at the initial points, but the blueshifts are finite and are
overcompensated by redshifts accumulated later along the ray. At the
intersection of a light cone with the SCS, the redshift profile has no
recognizable mark.

In Sec. \ref{maxred}, the location of the extremum-redshift hypersurface (ERH)
in spacetime is determined and displayed. (Along some rays, the maximum of
$z(r)$ is followed by a minimum. The maxima and minima lie on the ERH.) In Sec.
\ref{dlz}, the relations $D_A(z)$ and $D_L(z)$ are displayed, where $D_A$ and
$D_L$ are the angular-diameter distance and luminosity distance from the central
observer. These relations become double- or even triple-valued in the
blueshift-generating region.

In Sec. \ref{timing}, the timing of the last scattering in the L--T model
discussed here is compared to that in the $\Lambda$CDM model. The time
difference is small and finds a neat intuitive explanation.

Section \ref{conclu} presents the final summary of the results.

\section{Basic formulae}\label{basform}

\setcounter{equation}{0}

This is a brief summary of basic facts about the L--T model, included here
mainly in order to define the notation and conventions. For extended expositions
see Refs. \cite{PlKr2006,Kras1997}.

The metric of the model considered here is
\begin{equation}\label{2.1}
{\rm d} s^2 = {\rm d} t^2 - \frac {{R_{,r}}^2}{1 + 2E(r)}{\rm d} r^2 -
R^2(t,r)({\rm d}\vartheta^2 + \sin^2\vartheta \, {\rm d}\varphi^2),
\end{equation}
where $E(r)$ is an arbitrary function, and $R(t, r)$ obeys
\begin{equation}\label{2.2}
{R_{,t}}^2 = 2E(r) + 2M(r) / R,
\end{equation}
with $M(r)$ being another arbitrary function.

In the present paper, only the case $E > 0$ will be considered. Then the
solution of (\ref{2.2}) is
\begin{eqnarray}\label{2.3}
R(t,r) &=& \frac M {2E} (\cosh \eta - 1), \nonumber \\
\sinh \eta - \eta &=& \frac {(2E)^{3/2}} M \left[t - t_B(r)\right].
\end{eqnarray}
The function $t_B(r)$ is constant in the model considered here. The
$r$-coordinate is chosen so that
\begin{equation}\label{2.4}
M = M_0 r^3,
\end{equation}
and the remaining freedom of rescaling $r$ is used to take $M_0 = 1$ in
numerical calculations. The pressure is zero, and the mass density is
\begin{equation}  \label{2.5}
\kappa \rho = \frac {2{M_{,r}}}{R^2R_{,r}}, \qquad \kappa \df \frac {8\pi G}
{c^2}.
\end{equation}

Past radial null geodesics obey the equation
\begin{equation}\label{2.6}
\dr t r = - \frac {R_{,r}} {\sqrt{1 + 2E(r)}}.
\end{equation}
The redshifts $z(r)$ of the light sources lying along this light cone obey
\cite{Bond1947}, \cite{PlKr2006}
\begin{equation}\label{2.7}
\frac 1 {1 + z}\ \dr z r = \left[ \frac {R_{,tr}} {\sqrt{1 + 2E}} \right]_{\rm
ng}.
\end{equation}

The luminosity distance between the central observer and the light sources lying
along $t_{\rm ng}(r)$ is
\begin{equation}\label{2.8}
D_L(z) = (1 + z)^2 R\left(t_{\rm ng}(r), r\right).
\end{equation}
The particular solution of (\ref{2.2}) considered in this paper arose from the
requirement that the $D_L(z)$ given above coincides with that calculated from
the $\Lambda$CDM model,
\begin{equation}\label{2.9}
D_L(z) = \frac {1 + z} {H_0} \int_0^z \frac {{\rm d} z'} {\sqrt{\Omega_m (1 +
z')^3 + \Omega_{\Lambda}}},
\end{equation}
where $H_0$ is the value of the Hubble parameter at the present time. It is
related to the Hubble constant \cite{Plan2013}
\begin{equation}\label{2.10}
{\cal H}_0 = 67.1\ {\rm km}/({\rm s} \times {\rm Mpc})
\end{equation}
by
\begin{equation}\label{2.11}
H_0 = {\cal H}_0 / c.
\end{equation}
The two dimensionless parameters
\begin{equation}\label{2.12}
\left(\Omega_m, \Omega_{\Lambda}\right) \df \frac 1 {3{H_0}^2}
\left.\left(\frac {8\pi G \rho_0} {c^2}, - \Lambda\right)\right|_{t = t_o}
\end{equation}
obey $\Omega_m + \Omega_{\Lambda} \equiv 1$; $\rho_0$ is the present mean mass
density in the Universe. Their values
\begin{equation}\label{2.13}
(\Omega_m, \Omega_{\Lambda}) = (0.32, 0.68)
\end{equation}
also come from observations \cite{Plan2013}.

The requirement that the two expressions for $D_L(z)$, (\ref{2.8}) and
(\ref{2.9}), are equal along the radial null geodesic reaching the central
observer by now determines the functions $E(r)$, $t = t_{\rm ng}(r)$ obeying
(\ref{2.6}) and $z(r)$ obeying (\ref{2.7}). They were numerically calculated in
Ref. \cite{Kras2014} (see Sec. \ref{lightcone}, Fig. \ref{Efullrange} for $E(r)$
and Fig. \ref{conefirst} for $t_{\rm ng}(r)$).

The apparent horizon (AH) of the central observer is a locus where $R$,
calculated along the null geodesic given by (\ref{2.6}), changes from increasing
to decreasing, i.e., where $(\dril {} r) R(t_{\rm ng}(r), r) = 0$. With $\Lambda
= 0$, the general equation implicitly determining the function $t(r)$ along the
AH is
\begin{equation}\label{2.14}
R(t,r) = 2M(r).
\end{equation}
In the model considered here, the values of $r$ and $z$ at the intersection of
the PCPO with the AH are \cite{Kras2014}\footnote{The numbers calculated for
this paper by Fortran 90 are all at double precision -- to minimise
misalignments in the graphs.}
\begin{eqnarray}
r_{\rm AH} &=& 0.3105427968086945, \label{2.15} \\
z_{\rm AH} &=& 1.582430687623614. \label{2.16}
\end{eqnarray}

Since we consider here an L--T model with constant $t_B$, light emitted at the
Big Bang (BB) will be infinitely redshifted, as in the Robertson -- Walker (RW)
models \cite{Szek1980}, \cite{HeLa1984}, \cite{PlKr2006}.

The numerical units used here were introduced in Ref. \cite{Kras2014b}. They are
the numerical length unit (NLU) = the numerical time unit (NTU) related to the
usual units by
\begin{eqnarray}\label{2.17}
1\ {\rm NTU} &=& 1\ {\rm NLU} = 3 \times 10^4\ {\rm Mpc} \nonumber \\
&=& 9.26 \times 10^{23}\ {\rm km} = 9.8 \times 10^{10}\ {\rm y}.
\end{eqnarray}
In the above,
\begin{equation}\label{2.18}
c \approx 3 \times 10^5 {\rm km/s}
\end{equation}
was taken for the speed of light, and the following values of the conversion
factors were used \cite{unitconver}:
\begin{eqnarray}\label{2.19}
1\ {\rm pc} &=& 3.086 \times 10^{13}\ {\rm km}, \nonumber \\
1\ {\rm y} &=& 3.156 \times 10^7\ {\rm s}
\end{eqnarray}
In these units
\begin{eqnarray}
H_0 &=& 6.71\ ({\rm NLU})^{-1}, \label{2.20} \\
T &=& 13.819 \times 10^9\ {\rm y} = 0.141\ {\rm NTU}, \label{2.21}
\end{eqnarray}
where $T$ is the age of the $\Lambda$CDM Universe \cite{Plan2013}.

The age in the model used here is somewhat different:
\begin{equation}\label{2.22}
T_{\rm model} = - t_B = 0.1329433206844743\ {\rm NTU}.
\end{equation}

The mass associated to $M_0 = 1$ NLU in (\ref{2.4}) is
\begin{equation}\label{2.23}
m_0 \approx 1.5 \times 10^{54}\ {\rm kg},
\end{equation}
but it will appear only via $M_0$. The value of the gravitational constant used
in numerical calculations is \cite{Grav2010}
\begin{equation}\label{2.24}
G = 6.674 \times 10^{-11}\ {\rm m}^3/({\rm kg} \times {\rm s}^2).
\end{equation}

\section{Useful formulae for numerical calculations}\label{useform}

\setcounter{equation}{0}

With $t_B =$ constant and $M$ given by (\ref{2.4}) we have \cite{PlKr2006}
\begin{equation}\label{3.1}
R,_r = \left(\frac 3 r - \frac {E,_r} E\right)R + \left(\frac 3 2 \frac {E,_r} E
- \frac 3 r\right) \left(t - t_B\right) R,_t.
\end{equation}

In order to avoid a permanent singularity at the center of symmetry, the
function $E$ must be of the form \cite{PlKr2006}
\begin{equation}\label{3.2}
2 E(r) = r^2 [- k + {\cal F}(r)],
\end{equation}
where ${\cal F}(r)$ must obey
\begin{equation}\label{3.3}
\lim_{r \to 0}{\cal F} = 0,
\end{equation}
and the constant $k$ was determined in Ref. \cite{Kras2014}:
\begin{equation}\label{3.4}
k = - 21.916458.
\end{equation}
Substituting (\ref{2.4}) and (\ref{3.2}) in (\ref{2.3}) we obtain
\begin{eqnarray}
R &=& \frac {M_0 r} {-k + {\cal F}} (\cosh \eta - 1), \label{3.5} \\
\sinh \eta - \eta &=& \frac {(-k + {\cal F})^{3/2}} {M_0} \left(t - t_B\right).
\label{3.6}
\end{eqnarray}
Equations (\ref{3.5}) -- (\ref{3.6}) determine $R(t, r)$. The function ${\cal
F}(r)$ was numerically determined in Ref. \cite{Kras2014}.

{}From (\ref{3.1}) and (\ref{2.7}), using (\ref{2.2}) and (\ref{3.2}), we find:
\begin{eqnarray}\label{3.7}
\dr z r &=& \frac {1 + z} {\sqrt{1 + r^2 (-k + {\cal F})}} \nonumber  \\
&\times& \left\{\left[1 + \frac {r {\cal F},_r} {2 (-k + {\cal F})}\right]
\sqrt{- k + {\cal F}}\ \sqrt{\frac {\cosh \eta + 1} {\cosh \eta - 1}}\right.
\nonumber \\
&&\ \ \ \  - \left.\frac 3 2\ \frac {r {\cal F},_r (- k + {\cal F})} {M_0
(\cosh \eta - 1)^2}\ \left(t - t_B\right)\right\}.
\end{eqnarray}
The limit $r \to 0$ of this is, using (\ref{3.3}) and (\ref{3.6}),
\begin{equation}\label{3.8}
\dr z r(0) = \sqrt{-k}\ \left.\sqrt{\frac {\cosh \eta + 1} {\cosh \eta -
1}}\right|_{r = 0},
\end{equation}
where $\eta_0 \df \left.\eta\right|_{r = 0}$ is found by solving (\ref{3.6}) at
$r = 0$.

Since $t_B$ is given by (\ref{2.22}), and the function ${\cal F}(r)$ is given as
a numerical table, the solution of (\ref{2.14}) can be numerically found in the
form $t = t_{\rm AH}(r)$ from (\ref{3.6}), with $\eta(r)$ along the AH being
found from (\ref{3.5}) as follows:
\begin{equation}\label{3.9}
\eta_{\rm AH}(r) = \ln \left[x(r) + \sqrt{x^2(r) - 1}\right],
\end{equation}
where
\begin{equation}\label{3.10}
x(r) \df 1 + 2 r^2 (-k + {\cal F}).
\end{equation}

Substituting (\ref{3.2}), (\ref{3.5}) and (\ref{3.6}) in (\ref{3.1}) we obtain
\begin{eqnarray}\label{3.11}
R,_r &=& \left(1 - \frac {r {\cal F},_r} {-k + {\cal F}}\right) \frac {M_0} {-k
+ {\cal F}} (\cosh \eta - 1) \nonumber \\
&+& \frac 3 2\ \frac {r {\cal F},_r} {\sqrt{-k + {\cal F}}} \left(t - t_B\right)
\sqrt{\frac {\cosh \eta + 1} {\cosh \eta - 1}}.
\end{eqnarray}
The limit $r \to 0$ of this is
\begin{equation}\label{3.12}
R,_r(t, 0) = \frac {M_0} {(-k)} (\cosh \eta_0 - 1).
\end{equation}

\section{The past light cone of the present central observer}\label{lightcone}

\setcounter{equation}{0} \setcounter{table}{0}

To calculate the PCPO, $z(r)$ had to be calculated first. Since $z \to \infty$
at the BB, the numerical calculation broke down at $r = r_{\rm max}$, with $z =
z_{\rm max}$, where \cite{Kras2014}
\begin{equation}\label{4.1}
\left(\begin{array}{ll}
r_{\rm max} \\
z_{\rm max} \\
\end{array}\right) = \left(\begin{array}{ll}
1.045516839812362 \\
9.1148372886058313 \times 10^{225} \\
\end{array}\right).
\end{equation}

The model extends from the center of symmetry up to that flow line, at which the
PCPO reaches the BB. In practice this is the flow line corresponding to the
$r_{\rm max}$ given above. For $r > r_{\rm max}$, the model is not determined.
Extensions into the range $r > r_{\rm max}$ are possible, but are not
constrained by (\ref{2.9}), and are not considered here.

\begin{figure}[h]
\hspace{-1cm}
\includegraphics[scale=0.8]{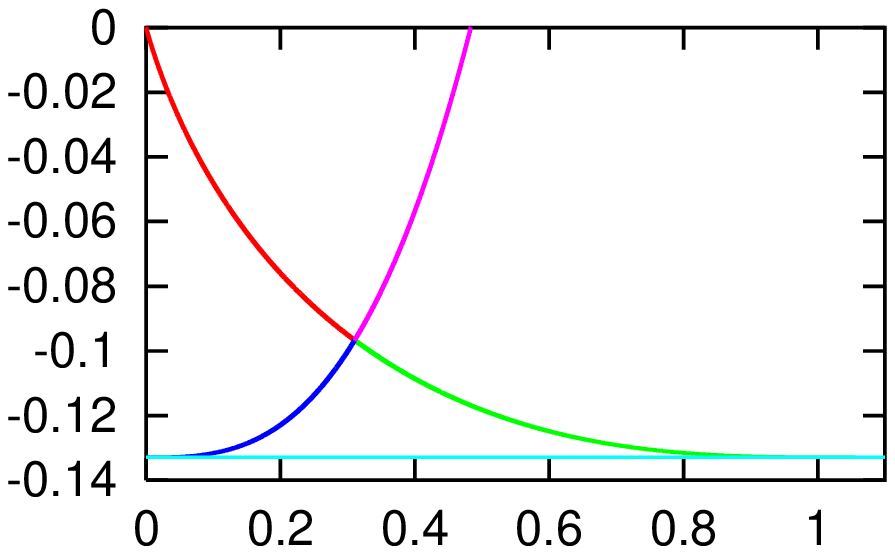}
${ }$ \\[-4cm]
\hspace{2.4cm}
\includegraphics[scale=0.35]{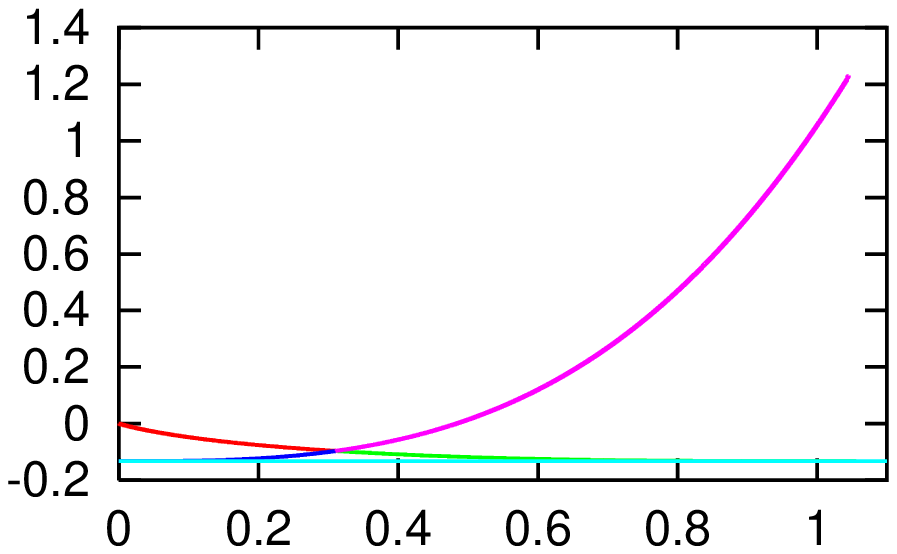}
\vspace{2cm} \caption{The profiles $t(r)$ of the past light cone of the present
central observer (the decreasing curve) and of the apparent horizon (the
increasing curve). The horizontal line is $t = t_B$. The inset shows the AH in
the full range of $r$. } \label{conefirst}
\end{figure}

The numerically calculated profile $t(r)$ of the PCPO is reproduced in Fig.
\ref{conefirst} (from Ref. \cite{Kras2014}), together with the profile of the AH
calculated from (\ref{3.6}) and (\ref{3.9}) -- (\ref{3.10}). The two profiles
intersect at the point $(r, t) = (r_{\rm AH}, t_{\rm AH})$, where $r_{\rm AH}$
is given by (\ref{2.15}) and \cite{Kras2014}
\begin{equation}\label{4.2}
t_{\rm AH} = -0.0966669255756665\ {\rm NTU}.
\end{equation}

The inset in Fig. \ref{conefirst} shows the AH up to the edge of the model. The
last point on it has the coordinates
\begin{equation}\label{4.3}
\left(\begin{array}{ll}
r_{\rm edge} \\
t_{\rm edge} \\
\end{array}\right) = \left(\begin{array}{ll}
1.0455189100976430 \\
1.2308128894377963 \\
\end{array}\right).
\end{equation}
This explains why the AH will not be seen in the graphs corresponding to late
times (see next sections).

\section{The shell crossings}\label{locateSCS}

\setcounter{equation}{0}

\begin{figure}[h]
\includegraphics[scale=0.6]{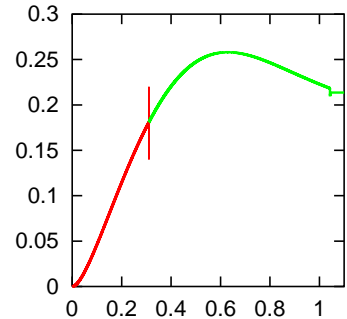}
\caption{The function $E(r)$ in the present model. The irregular segment at the
right end is caused by numerical fluctuations in the neighbourhood of the Big
Bang. The vertical stroke is at $r = r_{\rm AH}$. Since $E(r)$ becomes
decreasing for $r > r_{\rm sc}$, given by (\ref{5.1}), there are shell crossings
in that region. } \label{Efullrange}
\end{figure}

The shell crossing is a locus, where $R,_r = 0$ while $M,_r \neq 0$. As seen
from (\ref{2.5}), this is a curvature singularity, at which the mass density
becomes infinite, and changes sign if $R,_r$ does. When $t_B =$ constant and $E
> 0$, the necessary and sufficient condition for the absence of shell crossings
is $E,_r > 0$. However, the $E(r)$ calculated in Ref. \cite{Kras2014}, shown in
Fig. \ref{Efullrange}, has a maximum at
\begin{equation}\label{5.1}
r = r_{\rm sc} = 0.6293128978680214,
\end{equation}
and becomes decreasing for $r > r_{\rm sc}$. Thus, there are shell crossings in
the region $r > r_{\rm sc}$. The location of the SCS was not determined in Ref.
\cite{Kras2014}, and we shall do it here.

The equation $R,_r = 0$, using (\ref{3.11}), can be written as
\begin{equation}\label{5.2}
\frac 1 r + \frac {{\cal F},_r} {- k + {\cal F}}\ \left[\tfrac 3 2\ Q(\eta) -
1\right] = 0,
\end{equation}
where
\begin{equation}\label{5.3}
Q(\eta) \df \frac {\sinh \eta - \eta} {\sinh^3 \eta}\ (\cosh \eta + 1)^2.
\end{equation}
The function $Q(\eta)$ has the following properties
\begin{eqnarray}
&& Q(0) = 2 /3, \qquad \lim_{\eta \to \infty} Q(\eta) = 1, \label{5.4}
\\
&& \dr Q {\eta} > 0 \quad {\rm for} \quad 0 < \eta < \infty. \label{5.5}
\end{eqnarray}
The proof of (\ref{5.5}) is given in the Appendix \ref{prove48}. Thus, $Q$ is
monotonic in its full range.

On the SCS, $r > r_{\rm sc}$ must hold. The calculation stops at the $r_{\rm
max}$ given by (\ref{4.1}) because ${\cal F}(r)$ is undetermined for $r > r_{\rm
max}$. Consequently, for every $r \in (r_{\rm sc}, r_{\rm max})$, (\ref{5.2})
uniquely determines $\eta(r)$. Then, $t(r)$ along the SCS is calculated from
(\ref{2.3}):
\begin{equation}\label{5.6}
t(r) = t_B + \frac {M_0} {(- k + {\cal F})^{3/2}}\ [\sinh \eta(r) - \eta(r)].
\end{equation}

{}From (\ref{5.2}) one can see that at $r = r_{\rm sc}$, where $E,_r = 0$, we
have $Q = 1$, i.e. $\eta \to \infty$. Thus, the points of the SCS, at which $r =
r_{\rm sc}$, lie in the infinite future.

\begin{figure}[h]
\hspace{-3cm}
\includegraphics[scale=0.9]{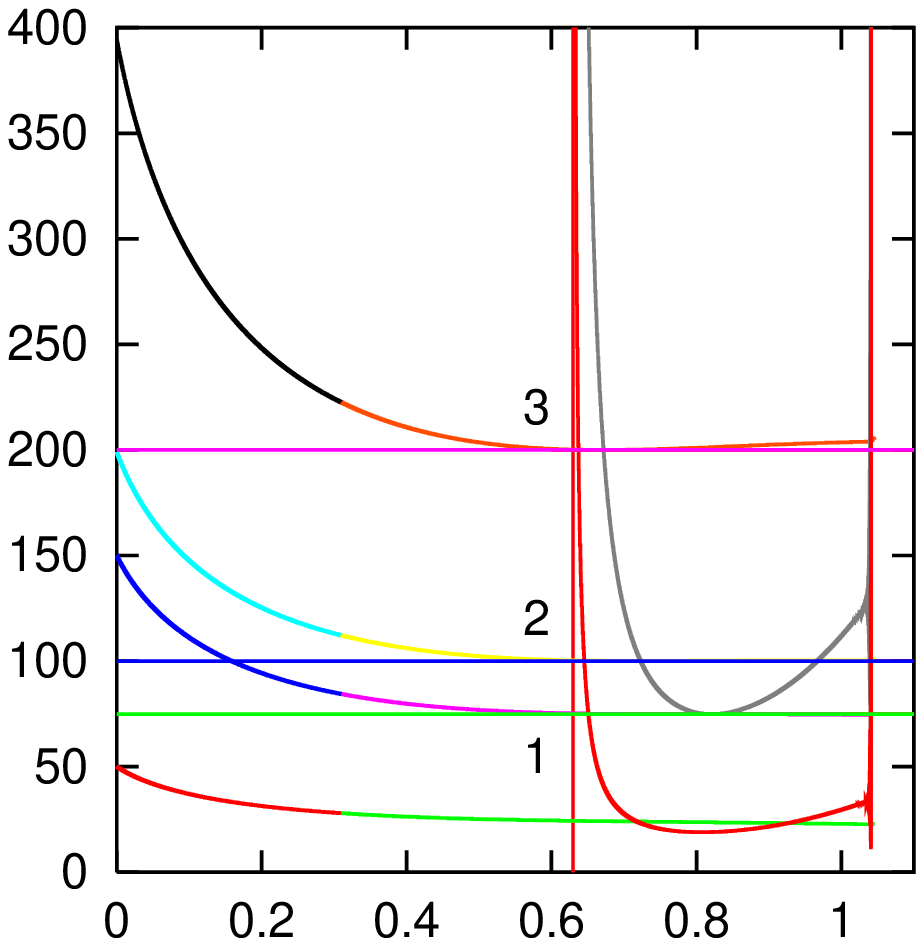}
${ }$ \\[-8.5cm]
\hspace{-1.3cm}
\includegraphics[scale=0.35]{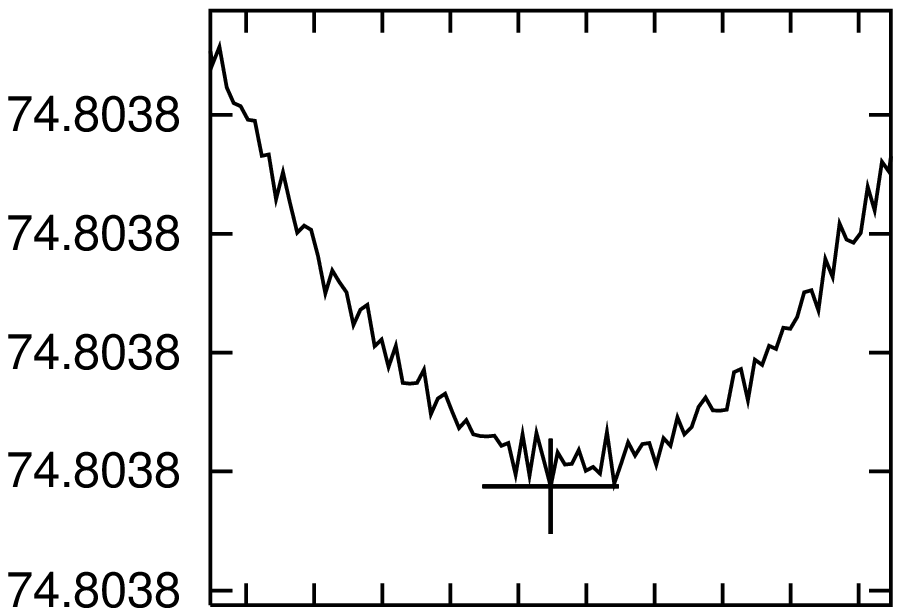}
\vspace{5.5cm}
\caption{{\bf Main panel:} See text for explanations. {\bf Inset:} The
neighbourhood of $r = r_{\rm min}$ on the shell-crossing set. The cross marks
the point $(r, t) = (r_{\rm min}, t_{\rm min})$ given by (\ref{5.7}) and
(\ref{5.8}). The leftmost tic on the $r$-axis is at $r = 0.82076$, the rightmost
one is at $r = 0.82094$, the tics are separated by $\Delta r = 2 \times
10^{-5}$.} \label{shellcross}
\end{figure}

The profile of the SCS given by (\ref{5.6}) is the upper ${\cal U}$-shaped curve
in the right part of Fig. \ref{shellcross}. The $t(r)$ on it decreases from $t =
+\infty$ at $r = r_{\rm sc}$ to
\begin{equation}\label{5.7}
t = t_{\rm min} = 74.803824384008095\ {\rm NTU}
\end{equation}
achieved at
\begin{equation}\label{5.8}
r = r_{\rm min} = 0.82084948116793433,
\end{equation}
and then increases with increasing $r$ up to the edge of the model given by
(\ref{4.1}). However, the numbers given in (\ref{5.7}) and (\ref{5.8}), which
emerged while calculating the $t(r)$ from (\ref{5.6}), are meaningful only up to
the fourth decimal digit because of numerical fluctuations in the SCS profile.
They are shown, in the vicinity of $r = r_{\rm min}$, in the inset in Fig.
\ref{shellcross}. The SCS has very large fluctuations for $r \to r_{\rm max}$,
seen at the right margin of Fig. \ref{shellcross}.

Figure \ref{shellcross} shows the intersections of various hypersurfaces with a
fixed 2-space of constant $\vartheta$ and $\varphi$. To get an idea about
spatial relations, one should imagine Fig. \ref{shellcross} being rotated around
the $r = 0$ axis -- this would be the intersections of those hypersurfaces with
the 3-space $\vartheta = \pi/2$.

The main panel of Fig. \ref{shellcross} shows several other curves that will be
introduced in later sections. The lower ${\cal U}$-shaped curve is the
extremum-redshift profile given by (\ref{10.2}) -- (\ref{10.3}). The lowest of
the other four curves is the profile of the light cone that hits the central
observer at $t = 50$ NTU; it will become clear in Sec. \ref{redprof} why it is
special. The remaining three curves are the profiles of the past light cones of
the central observer that intersect the SCS at the times $t_{\rm min}$, $t_2$
and $t_3$ given by (\ref{5.7}), (\ref{5.10}) and (\ref{5.12}). Note that all the
light cones that intersect the SCS are horizontal at the intersection points. At
the scale of this figure, the light cone from Fig. \ref{conefirst} seems to
coincide with the $r$-axis. The left vertical line marks $r = r_{\rm sc}$, the
right one is an artifact of numerical fluctuations as $r \to r_{\rm max}$. The
horizontal lines 1, 2 and 3 mark the characteristic times explained below.

The time given by (\ref{5.7}) can be written as
\begin{equation}\label{5.9}
t_{\rm min} \approx 562.4 T,
\end{equation}
where $T$ is given by (\ref{2.21}). Thus, since the SCS lies far to the future
of the light cone from Fig. \ref{conefirst}, it has no influence on the past and
present observations of the central observer. It will influence her observations
beginning at $t = t_F$, where $t_F$ is that instant, at which the light ray
issued at $(r, t) = (r_{\rm min}, t_{\rm min})$ hits the center $r = 0$. The
$t_F$ will be determined in Sec. \ref{SCSinflu}.

The horizontal lines marked ``1'', ``2'' and ``3'' in Fig. \ref{shellcross} are
at those values of $t$, at which the profiles of density will be calculated in
Sec. \ref{densities}. The line marked ``1'', is tangent to the SCS at $r =
r_{\rm min}$. The line marked ``2'', at
\begin{equation}\label{5.10}
t = t_2 = 100\ {\rm NTU},
\end{equation}
has two points of intersection with the SCS profile. The positions of those
points can be read off from the table of values of $t(r)$ along the SCS, they
are
\begin{equation}\label{5.11}
r = r_{2a} \approx 0.722829 \quad {\rm and} \quad r = r_{2b} \approx 0.967.
\end{equation}
The line marked ``3'' intersects the SCS at one point, with the coordinates
\begin{equation}\label{5.12}
t = t_3 = 200\ {\rm NTU}, \quad r = r_3 \approx 0.671732.
\end{equation}

\section{Density distributions on hypersurfaces of constant
$t$}\label{densities}

\setcounter{equation}{0}

We will now calculate the density distributions in our model along a few
characteristic hypersurfaces. The curves in all the figures in this section have
large numerical fluctuations at $r \to r_{\rm max}$ that look like vertical
bars, these are ignored in the captions and explanations.

\begin{figure}[h]
\includegraphics[scale=0.85]{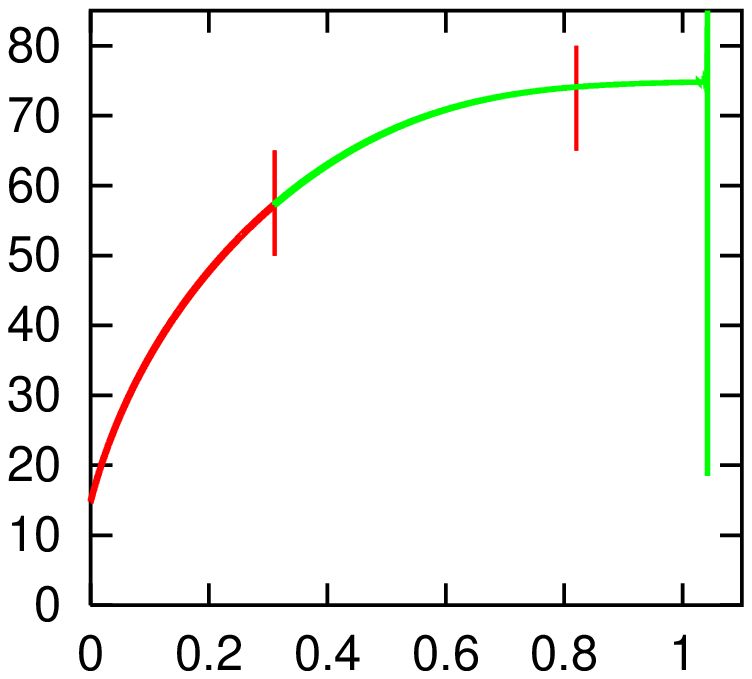}
${ }$ \\[-4.5cm]
\hspace{1.8cm}
\includegraphics[scale=0.45]{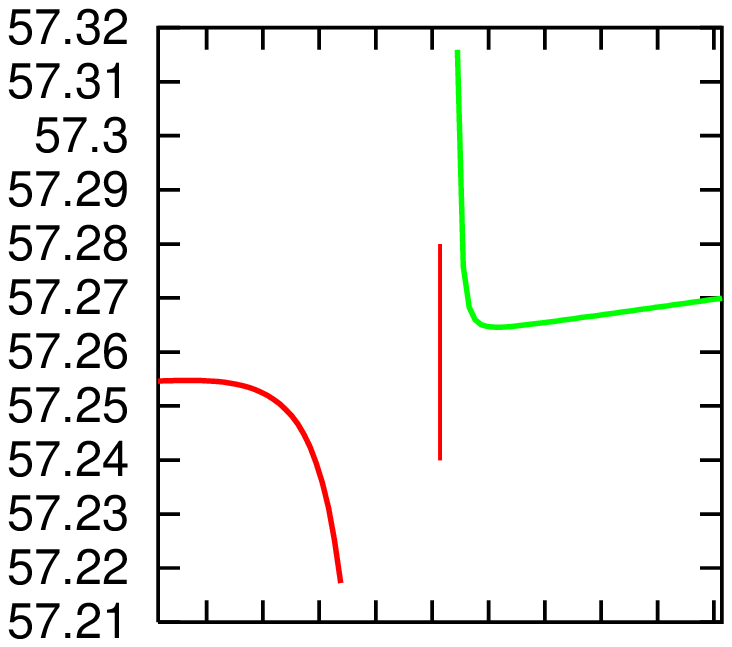}
\vspace{5mm}
\caption{{\bf Main panel:} The mass density (\ref{2.5}) calculated along the
hypersurface $t = 0$ as a function of $r$. The quantity on the vertical axis is
$\kappa \rho$, measured in $({\rm NLU})^{-2}$. The left vertical bar marks $r =
r_{\rm AH}$, the right one marks $r = r_{\rm sc}$. {\bf Inset:} Enlarged view of
the neighbourhood of $r = r_{\rm AH}$. The leftmost tic on the horizontal axis
is at $r = 0.31046$, the rightmost tic is at $r = 0.31064$, the tics are
separated by $\Delta r = 2 \times 10^{-5}$. The discontinuity is a consequence
of numerical errors. } \label{densityatt0}
\end{figure}

We first calculate the density at the time $t = 0$, where the light cone of Fig.
\ref{conefirst} has its vertex. To this end, we use (\ref{2.2}), (\ref{2.4}),
(\ref{3.2}), (\ref{2.22}) and $t = 0$ in (\ref{3.5}) -- (\ref{3.6}) and in
(\ref{3.1}) to calculate $R(0, r)$ and $R,_r(0, r)$, then we substitute the
results in (\ref{2.5}). The resulting graph of $\kappa \rho(0, r)$ is shown in
the main panel of Fig. \ref{densityatt0}. This is a void profile, in agreement
with the conventional wisdom,\footnote{Many authors just reflexively use the
term ``void models'' to denote L--T models mimicking accelerated expansion, see
references in Refs. \cite{CBKr2010,BCKr2011}. In general, this is not correct --
as demonstrated by the models of Refs. \cite{CBKr2010} and \cite{Kras2014}.} but
it has a cusp rather than a smooth minimum at the center.\footnote{See footnote
\ref{fnote1}.} The hypersurface $t = 0$ does not intersect the SCS, so this
density is finite in the whole range. The inset in Fig. \ref{densityatt0} shows
an enlarged view of the neighbourhood of $r = r_{\rm AH}$. The curve $\kappa
\rho (0, r)$ has a discontinuity at $r = r_{\rm AH}$ caused by numerical errors
in calculating $R$ and $R,_r$. These errors are consequences of inaccuracy in
calculating $E(r)$ in the neighbourhood of $r = r_{\rm AH}$, reported in Ref.
\cite{Kras2014}.

The central density in this profile is
\begin{equation}\label{6.1}
\kappa \rho(0, 0) = 14.837453949082986\ ({\rm NLU})^{-2},
\end{equation}
which corresponds to
\begin{equation}\label{6.2}
\rho(0, 0) \approx 0.93 \times 10^{-30}\ {\rm g/cm}^3.
\end{equation}
This is smaller than the present density in the $\Lambda$CDM model, which can be
calculated\footnote{The conversion to g/cm$^3$ in (\ref{6.1}) -- (\ref{6.7}) was
done using (\ref{2.18}) for the velocity of light and (\ref{2.24}) for the
gravitational constant.} from (\ref{2.12}) using (\ref{2.10}), (\ref{2.11}) and
(\ref{2.13}):
\begin{eqnarray}
\kappa \rho_0 &\approx& 43.075\ ({\rm NLU})^{-2}, \label{6.3} \\
\rho_0 &\approx& 2.7 \times 10^{-30} {\rm g/cm}^3. \label{6.4}
\end{eqnarray}

\begin{figure}[h]
\includegraphics[scale=0.85]{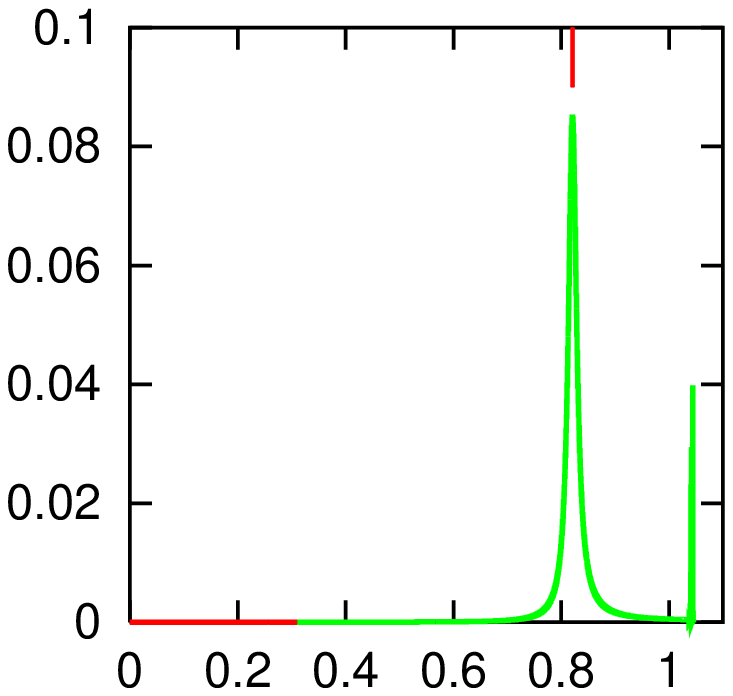}
${ }$ \\[-5.5cm]
\hspace{-5mm}
\includegraphics[scale=0.4]{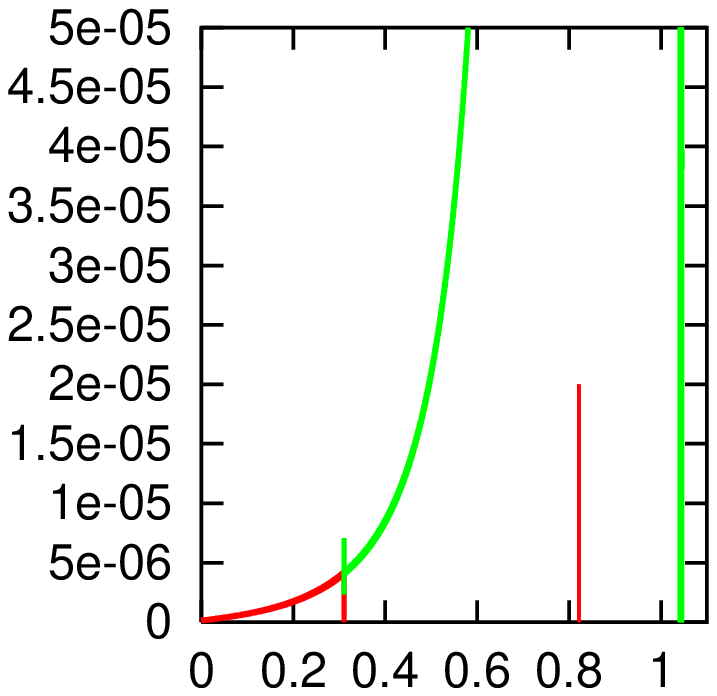}
\vspace{25mm}
 \caption{{\bf Main panel:} The mass density (\ref{2.5}) calculated along the
hypersurface $t = t_{\rm min}$, given by (\ref{5.7}), as a function of $r$. The
units on the vertical axis are the same as in Fig. \ref{densityatt0}. The
vertical bar at the top marks the value $r = r_{\rm min}$, given by (\ref{5.8}).
{\bf Inset:} Enlarged view of the neighbourhood of the $r$-axis. The right
vertical bar marks $r = r_{\rm min}$. What looks like another bar at $r = r_{\rm
AH}$ is a numerical fluctuation. } \label{densityatt1}
\end{figure}

Figure \ref{densityatt1} shows the graph of $\kappa \rho(t_{\rm min}, r)$, where
$t_{\rm min}$ is the time, given by (\ref{5.7}), at which the hypersurface $t =
t_{\rm min}$ is tangent to the SCS at the $r = r_{\rm min}$, given by
(\ref{5.8}). The density goes to infinity there, but stays positive for  $r >
r_{\rm min}$. The main panel shows the position of the peak in
density,\footnote{Because of numerical inaccuracies, no point on the line $t =
t_{\rm min}$ actually coincides with any point on the SCS. Therefore, at $r =
r_{\rm min}$, the numerically calculated density becomes much larger than
elsewhere, but is still finite.} the inset shows the values of $\kappa \rho$ in
a vicinity of the $r$-axis. The central density is
\begin{eqnarray}\label{6.5}
\kappa \rho(t_{\rm min},0) &=& 1.39237296085099607 \times 10^{-7} ({\rm
NLU})^{-2} \nonumber \\
\Longrightarrow \rho(t_{\rm min},0) &\approx& 8.73 \times 10^{-39}\ {\rm g/cm}^3.
\end{eqnarray}

\begin{figure}[h]
\hspace{-1.4cm}
\includegraphics{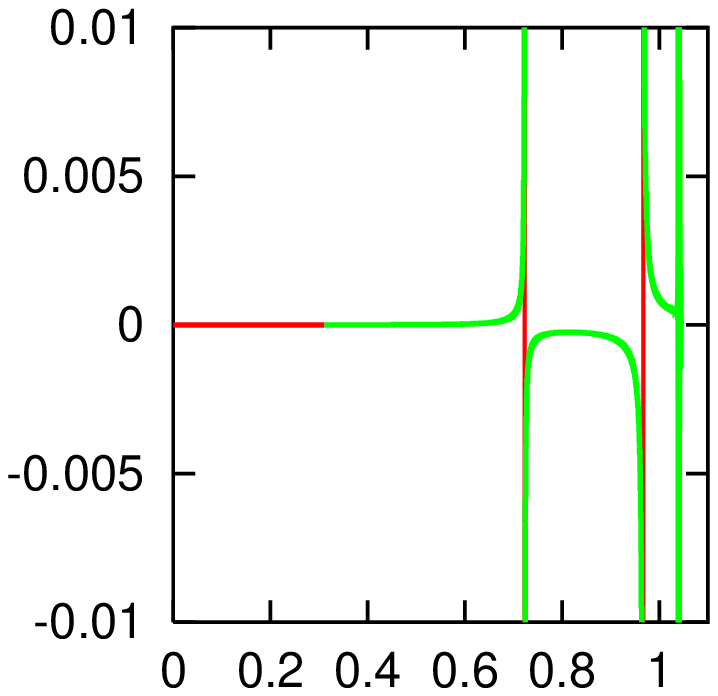}
${ }$ \\[-4cm]
\hspace{-2cm}
\includegraphics[scale=0.35]{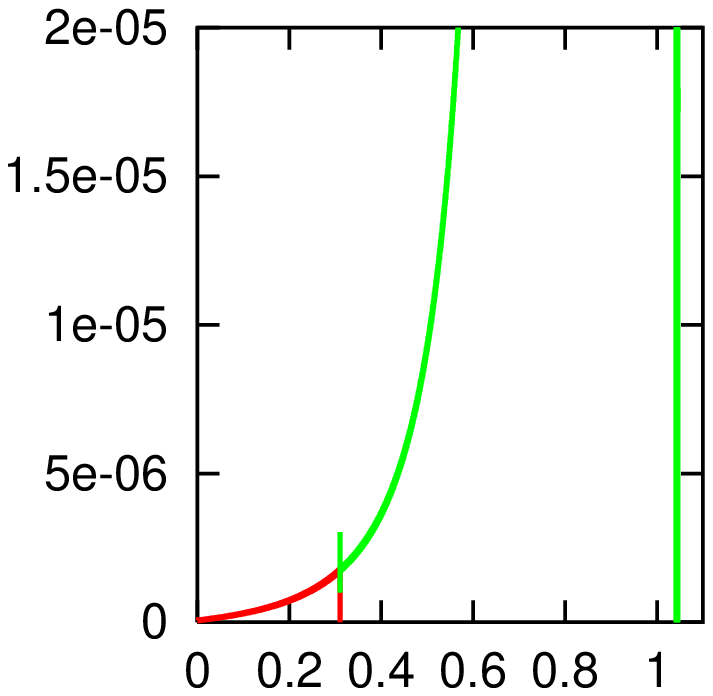}
 \vspace{1cm}
\caption{{\bf Main panel:} The mass density (\ref{2.5}) as a function of $r$,
calculated along the hypersurface $t = t_2 = 100$ NTU. The units on the
vertical axis are the same as in Fig. \ref{densityatt0}. The vertical bars mark
the values $r = r_{2a}$ and $r = r_{2b}$. {\bf Inset:} Enlarged view of the
neighbourhood of the $r$-axis. What looks like a vertical bar is a numerical
fluctuation at $r = r_{\rm AH}$. } \label{densityatt2}
\end{figure}

Figure \ref{densityatt2} shows the graph of $\kappa \rho(t_2, r)$, where $t_2 =
100$ NTU. The hypersurface $t = t_2$ intersects the SCS at two values of $r$,
given by (\ref{5.11}). At $r \to {r_{2a}}_-$, the density goes to $+ \infty$ and
becomes negative for $r > r_{2a}$. At $r \to {r_{2b}}_-$, the density goes to $-
\infty$ and becomes positive for $r > r_{2b}$. The main panel shows the peaks,
the inset shows $\kappa \rho$ in a vicinity of the $r$-axis. The central density
is
\begin{eqnarray}\label{6.6}
\kappa \rho(t_2,0) &=& 5.83260001345375805 \times 10^{-8} ({\rm NLU})^{-2}
\nonumber \\
\Longrightarrow \rho(t_2,0) &\approx& 3.66 \times 10^{-39}\ {\rm g/cm}^3.
\end{eqnarray}

\begin{figure}[h]
\hspace{-1cm}
\includegraphics[scale=0.95]{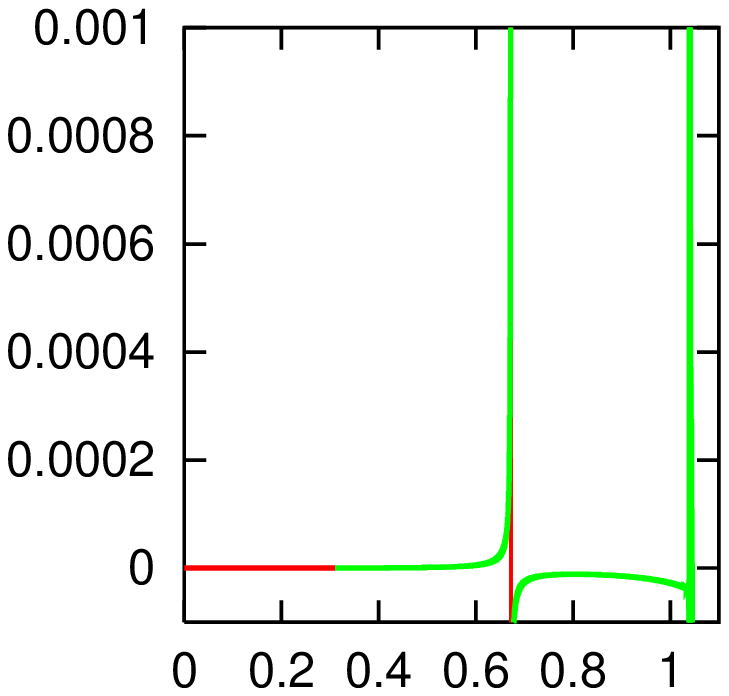}
${ }$ \\[-6cm]
\hspace{-1.6cm}
\includegraphics[scale=0.35]{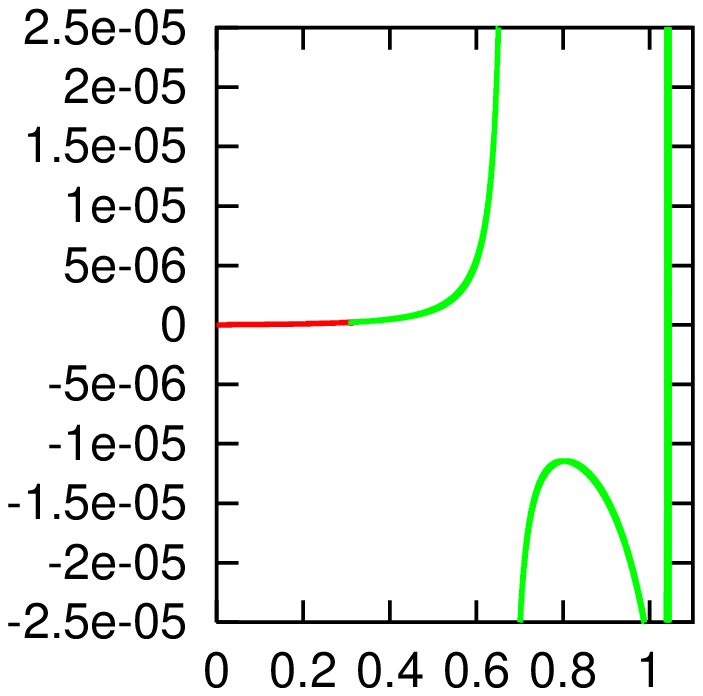}
 \vspace{3cm}
\caption{{\bf Main panel:} The mass density (\ref{2.5}) as a function of $r$,
calculated along the hypersurface $t = t_3 = 200$ NTU, which intersects the SCS
at the value of $r$ given by (\ref{5.12}). The units on the vertical axis are
the same as in Fig. \ref{densityatt0}. The vertical bar marks the value $r =
r_3$. {\bf Inset:} Enlarged view of the neighbourhood of the $r$-axis. }
\label{densityatt3}
\end{figure}

Figure \ref{densityatt3} shows the graph of $\kappa \rho(t_3, r)$, where $t_3 =
200$ NTU. The hypersurface $t = t_3$ intersects the SCS at $r = r_3$, given by
(\ref{5.12}). At $r \to {r_3}_-$, the density goes to $+ \infty$ and becomes
negative for $r > r_3$. The main panel shows the position of the peak in
density, the inset shows $\kappa \rho$ in a vicinity of the $r$-axis. The
central density is
\begin{eqnarray}\label{6.7}
\kappa \rho(t_3,0) &=& 7.29952961986280660 \times 10^{-9} ({\rm NLU})^{-2}
\nonumber \\
\Longrightarrow \rho(t_3,0) &\approx& 4.58 \times 10^{-40}\ {\rm g/cm}^3.
\end{eqnarray}

The transitions between the situations shown in Figs. \ref{densityatt0} --
\ref{densityatt3} occur in a nearly continuous way. At $t < t_{\rm min}$, there
is no singularity in $\rho(t, r)$. As $t \to t_{{\rm min} -}$, the density at $r
= r_{\rm min}$ goes to infinity, but stays positive on both sides of $r_{\rm
min}$. When $t$ increases above $t_{\rm min}$, the two branches of $\rho(t, r)$
that go to $+ \infty$ move sideways away from $r_{\rm min}$, and a third branch
appears that goes to $- \infty$ at $r \to r_{2a+}$ and $r \to r_{2b-}$. The
situation in Fig. \ref{densityatt3} does not really differ from that in Fig.
\ref{densityatt2}; the second infinity in $\rho$ has simply moved out of the
region covered by our model.

\section{The density distribution along the past light cone of the present
observer}\label{densoncone}

\setcounter{equation}{0}

\begin{figure}[h]
\includegraphics{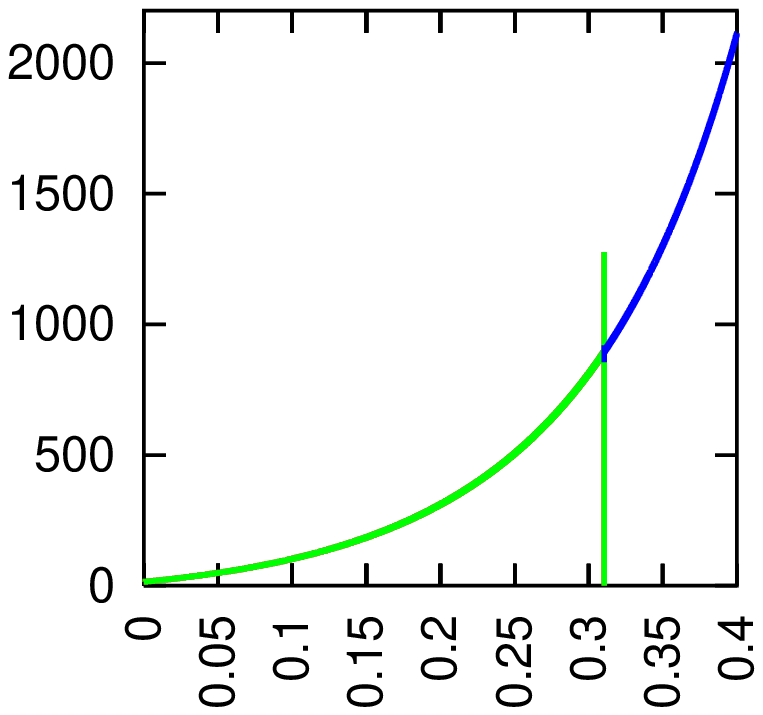}
${ }$ \\[-7.3cm]
\hspace{-2mm}
\includegraphics[scale=0.5]{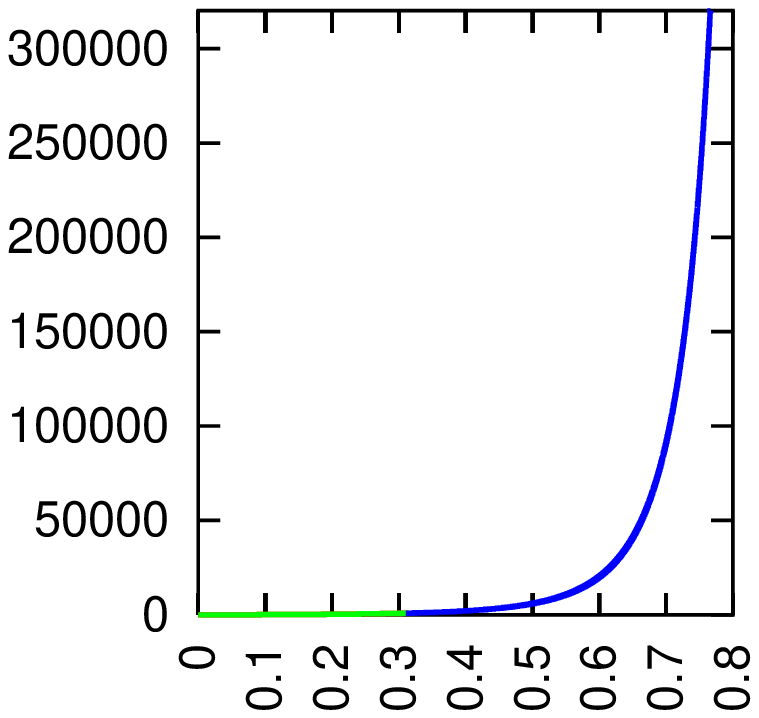}
\vspace{3cm}
\hspace{1cm}
\includegraphics[scale=0.45]{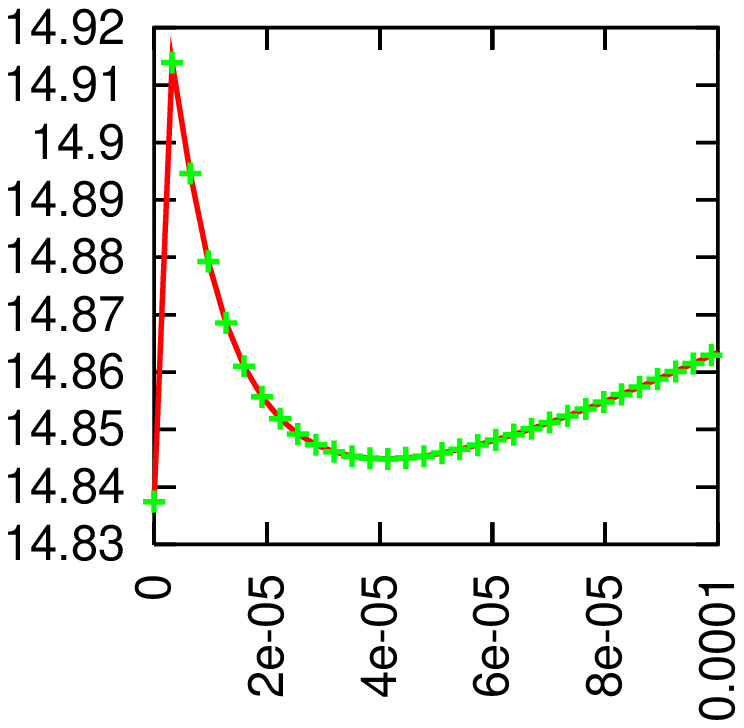}
\includegraphics[scale=0.45]{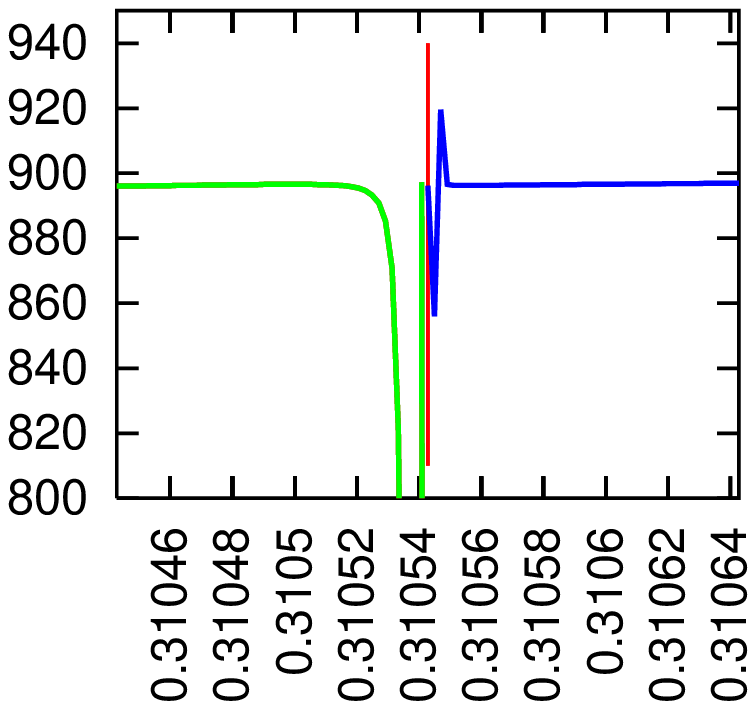}
\caption{The mass density (\ref{2.5}) calculated along the light cone of Fig.
\ref{conefirst} as a function of $r$. The units of $\kappa \rho$ are the same as
in Fig. \ref{densityatt0}. See text for explanations.} \label{densitiesofr}
\end{figure}

The hypersurface $t = 0$ contains events simultaneous with $(t, r) = (0, 0)$, so
the density distribution on it is unobservable at $(0, 0)$. The observable
quantity is the density along the observer's past light cone. It was calculated
from (\ref{2.5}), where (\ref{2.4}) was used for $M(r)$ and the numerical table
for $R(t_{\rm ng}(r), r)$ was taken from Ref. \cite{Kras2014}. In calculating
$R,_r(t_{\rm ng}(r), r)$ from (\ref{3.1}), eq. (\ref{3.2}) was used for $E$, eq.
(\ref{2.2}) was used for $R,_t$, and the numerical tables for ${\cal F}(r)$ and
$t_{\rm ng}(r)$ were taken from Ref. \cite{Kras2014}. In order to test the
numerical accuracy, the calculation of $\kappa \rho(t_{\rm ng}(r), r)$ in the
range $r \in [0, r_{\rm AH}]$ was carried out in two ways: forward from $r = 0$
and backward from $r = r_{\rm AH}$, using the relevant tables for $t_{\rm
ng}(r)$ from Ref. \cite{Kras2014}. Figure \ref{densitiesofr} shows $\kappa
\rho(t_{\rm ng}(r), r)$ in four separate views.

The main panel shows the graph of $\kappa \rho$ in the range $r \in [0,0.4]$. At
$r = 0$, the value of $\kappa \rho$ is, of course, the same as in (\ref{6.1}).
What looks like a vertical bar is a numerical fluctuation at $r = r_{\rm AH}$.
The curves calculated forward from $r = 0$ and backward from $r = r_{\rm AH}$
coincide perfectly, not only at this scale (see below).

The inset in the main panel shows the same graph in the range $r \in [0,0.8]$.
With $r$ approaching the $r_{\rm max}$ given by (\ref{4.1}), $\rho(t_{\rm
ng}(r), r)$ goes to infinity very fast. The numerical calculation broke down
already at
\begin{equation}\label{7.1}
r = r_{\rm bd} = 1.0296253299989211,
\end{equation}
with the largest value of $\kappa \rho$ yet calculated being
\begin{equation}\label{7.2}
\kappa \rho_{\rm bd} = 1.49073211697326822 \times 10^{18}\ ({\rm NLU})^{-2}.
\end{equation}

The lower left panel of Fig. \ref{densitiesofr} shows the neighbourhood of $r =
0$ and is meant to demonstrate that the curve calculated forward from $r = 0$
(the continuous line) agrees perfectly with the one calculated backward from $r
= r_{\rm AH}$ (the crosses) even at this scale. The fluctuation in the first
step is inherited from the numerical calculation of ${\cal F}(r)$ in Ref.
\cite{Kras2014}.

The lower right panel shows the neighbourhood of $r = r_{\rm AH}$ (marked with
the vertical bar), and the fluctuation, also inherited from ${\cal F}(r)$.
Except for the segment $r \in [r_{\rm AH} - \varepsilon, r_{\rm AH} +
\varepsilon]$, where $\varepsilon \approx 4 \times 10^{-5}$, the two parts of
the graph fit together satisfactorily.

\begin{figure}[h]
\includegraphics[scale=0.6]{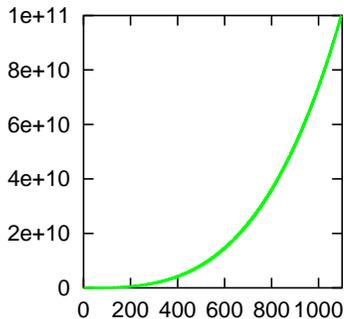}
\caption{The mass density (\ref{2.5}) along the light cone of Fig.
\ref{conefirst} as a function of $z$. The units of $\kappa \rho$ are the same as
in Fig. \ref{densityatt0}. } \label{densitiesofz}
\end{figure}

The numerical tables for $\kappa \rho(r)$ along the light cone, discussed above,
and for $z(r)$, calculated in Ref. \cite{Kras2014}, were then combined to
produce the table for $\kappa \rho(z)$. This does not differ in shape from
$\kappa \rho(r)$, except that $z \to \infty$ at the BB, so the horizontal axis
is stretched compared to that in Fig. \ref{densitiesofr}. The graph of $\kappa
\rho(z)$ is shown in Fig. \ref{densitiesofz}. The segment $z \in [0, z_{\rm
AH}]$, where $z_{\rm AH}$ is given by (\ref{2.16}), and the numerical
fluctuation at $z_{\rm AH}$, are squeezed to an invisible size. The right end of
the $z$-axis is at $z = 1100$, which is close to the redshift at last scattering
\cite{Luci2004},
\begin{equation}\label{7.3}
z_{\rm ls} = 1089.
\end{equation}

\section{The future observations of the shell crossings}\label{SCSinflu}

\setcounter{equation}{0}

As stated before, the SCS lies far to the future of the PCPO from Fig.
\ref{conefirst}. In order to find out, at what time the shell crossing will make
itself seen for the central observer, one must calculate the profile of that
past light cone of the central observer that is tangent to the SCS at $(r, t) =
(r_{\rm min}, t_{\rm min})$ given by (\ref{5.7}) -- (\ref{5.8}). This was done
by numerical fitting, and the result is the third curve from above in the left
part of Fig. \ref{shellcross}. It will reach the central observer at
\begin{equation}\label{8.1}
t_F \approx 149.965\ {\rm NTU} \approx 1063.58 T,
\end{equation}
where $T$ is given by (\ref{2.21}). This is the first instant, at which the
central observer will get a signal from the SCS. See Sec. \ref{redprof} for more
on this.

Since the equation of the SCS is $R,_r = 0$, Eq. (\ref{2.6}) implies that a
light ray intersecting the SCS must be horizontal in the $(t, r)$ coordinates at
the intersection point. The three uppermost rays in Fig. \ref{shellcross}
illustrate this.

Let us now compare the travel times of two rays: one that is emitted at the BB
and reaches the central observer at present $(t = 0)$, and one that is emitted
at the minimum of the SCS and reaches the central observer at $t = t_F$. The
first time is given by (\ref{2.22}), the second one is, from (\ref{5.7}) and
(\ref{8.1})
\begin{equation}\label{8.2}
t_F - t_{\rm min} \approx 75.161\ {\rm NTU}.
\end{equation}
Thus $(t_F - t_{\rm min})/T_{\rm model} \approx 565.12$, which may seem
surprisingly large. A ray emitted from the minimum of the SCS proceeds from
$r_{\rm min}$ to the observer at $r = 0$, which seems to be only part of the way
from $r_{\rm max}$ given by (\ref{4.1}) to $r = 0$, and yet the journey of the
later-emitted ray lasts much longer than the present age of the Universe. This
difference is a consequence of the illusion created by the comoving coordinates
that we are using throughout this paper. The values of $r$ are constant along
the flow lines of the cosmic fluid and are not measures of distance from the
symmetry center. Such a measure is, for example, $\int \frac {R,_r {\rm d} r}
{\sqrt{1 + 2E}}$ calculated at constant $t$. In consequence of expansion of the
Universe, a particle at $r = r_{\rm min}$ is farther from the center at $t =
t_{\rm min}$ than at any $t < t_{\rm min}$. Thus, the ray emitted from $r =
r_{\rm min}$ at the $t_{\rm min}$ given by (\ref{5.7}) has a much larger
distance to cover before reaching $r = 0$ than a ray emitted from the same
$r_{\rm min}$ at the BB, to reach the observer at $t = 0$.

\section{The redshift profiles along various light cones}\label{redprof}

\setcounter{equation}{0}

\begin{figure}[h]
\hspace{-2.3cm}
\includegraphics[scale=0.65]{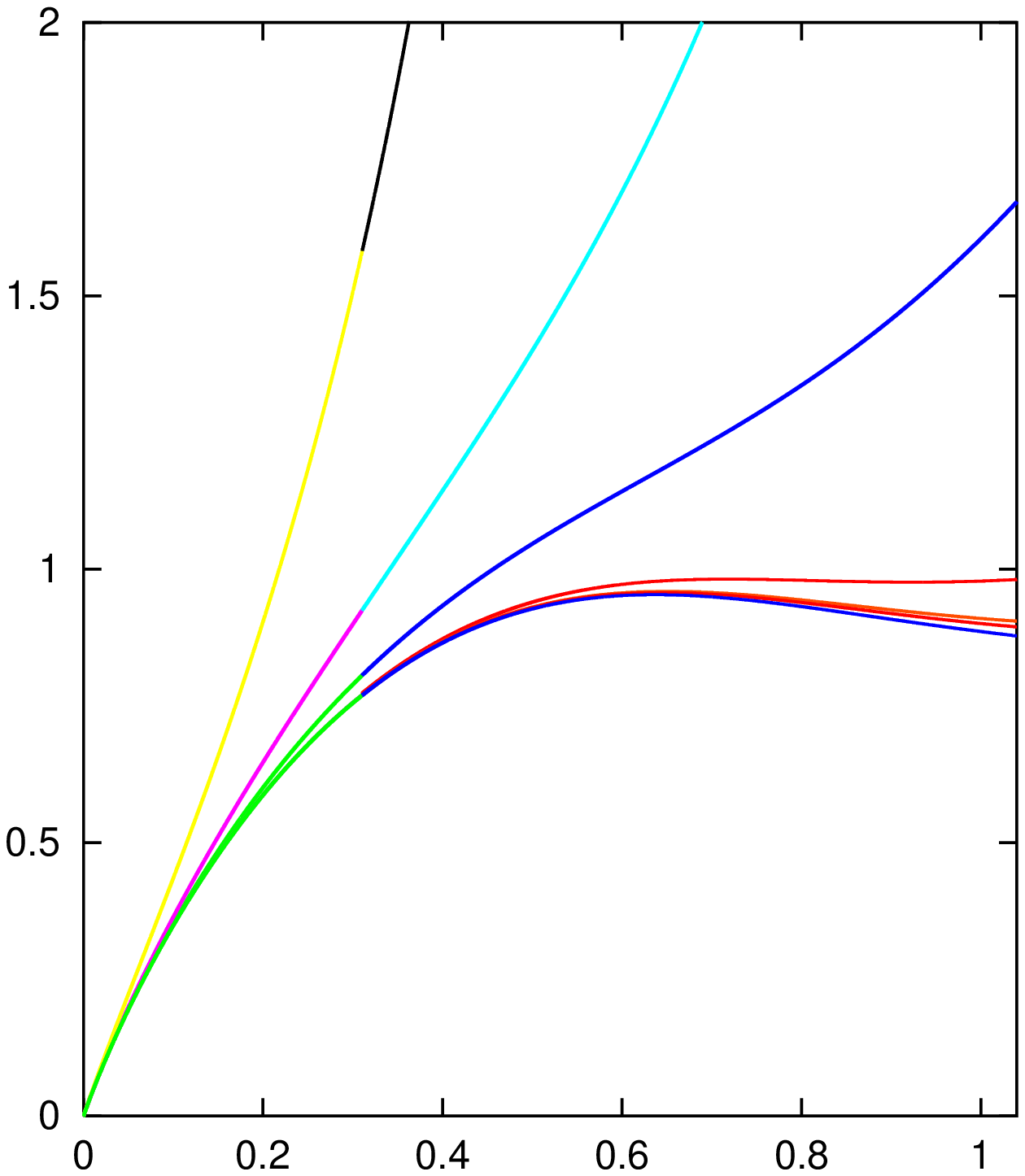}
${ }$ \\[-4.5cm]
\hspace{10.5cm}
\includegraphics[scale=0.6]{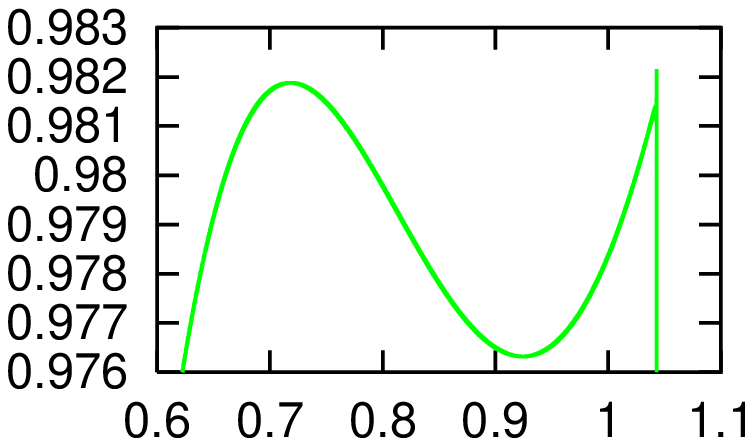}
\vspace{1cm}
\caption{Graphs of $z(r)$ along the rays reaching the central observer at
various times from $t = 0$ NTU to $t \approx 394.3$ NTU. See text for details.
{\bf Inset:} Enlarged view of the right part of curve \# 4 from the main graph.
It clearly shows that $z(r)$ along this ray has a local maximum and a local
minimum at $(r, z)$ given by (\ref{9.1}) and (\ref{9.2}).} \label{drawlatezall}
\end{figure}

It is interesting to trace the behaviour of redshift along various light rays.
The profiles of $z(r)$ shown in Fig. \ref{drawlatezall} were calculated by
solving (\ref{3.7}) along the rays reaching the central observer at the times
given below. The numbering begins at the upper edge of the figure from the left,
and continues along the right edge from top to bottom.

\begin{enumerate}
\item $t = 0$ NTU. This is $z(r)$ along the ray emitted at the BB that
    reaches the central observer at present. It was calculated in Ref.
    \cite{Kras2014} from the input data defining the model, and now
    re-calculated by solving (\ref{3.7}). Tables \ref{comprof1} and
    \ref{comprof2} show the precision with which the two results
    (dis-)agree. For Table \ref{comprof2} the initial values of $r$ and $z$
    are given by (\ref{2.15}) and (\ref{2.16}). Wherever the difference is
    nonzero, the now-calculated $z(r)$ is greater.

\item $t = 1$ NTU.

\item $t = 5$ NTU.

\item $t = 50$ NTU. This ray and the next three are shown in Fig.
\ref{shellcross}.

\item $t \approx 149.965$ NTU. This is $z(r)$ along the ray emitted from the
SCS at $(r, t) = (r_{\rm min}, t_{\rm min})$ given by (\ref{5.7}) --
(\ref{5.8}).

\item $t \approx 199.083$ NTU. This is $z(r)$ along the ray that intersects the
SCS at $t = t_2 = 100$ NTU.

\item $t \approx 394.314$ NTU. This is $z(r)$ along the ray that intersects the
SCS at $t = t_3 = 200$ NTU.
\end{enumerate}

\begin{table}[h]
\begin{center}
\caption{Differences between the two $z(r)$ functions on ray 1 when (\ref{3.7})
is integrated from $r = 0$}\label{comprof1}
\begin{tabular}{|l|l|}
 \hline \hline
At  & the difference is \\
 \hline \hline
$r$ close to 0 & 0 (perfect agreement \\
 \  & down to $\Delta r = 10^{-6}$) \\
 \hline
$r$ close to $r_{\rm AH}$ & $\Delta z = 1.2 \times 10^{-4}$ \\
 \hline
$r = 0.7$, where $z \approx 10.4$ &  $\Delta z = 2.7 \times 10^{-3}$ \\
 \hline
$r = 1.01$, where $z \approx 1050$ & $\Delta z \approx 7$ \\
 \hline \hline
\end{tabular}
\end{center}
\end{table}

\begin{table}[h]
\begin{center}
\caption{Differences between the two $z(r)$ functions on ray 1 when (\ref{3.7})
is integrated from $r = r_{\rm AH}$}\label{comprof2}
\begin{tabular}{|l|l|}
 \hline \hline
At  & the difference is \\
 \hline \hline
$r$ close to $r_{\rm AH}$ & 0 (perfect agreement \\
 \  & down to $\Delta r = 10^{-6}$) \\
 \hline
$r = 0.7$ &  $\Delta z = 2.2 \times 10^{-3}$ \\
 \hline
$r = 1.01$ & $\Delta z \approx 3.8$ \\
 \hline \hline
\end{tabular}
\end{center}
\end{table}

A few facts about the graphs in Fig. \ref{drawlatezall} are noteworthy:

1. On curves \# 1 to 3, $z(r)$ is monotonically increasing along the past light
cones.

2. On curve \# 4, whose corresponding light cone comes near the SCS, but does
not intersect it (see Fig. \ref{shellcross} -- this is the lowest of the four
light cone profiles shown there), $z(r)$ goes through a maximum and then through
a minimum before the ray escapes through the edge of the model. The coordinates
of these extrema are
\begin{eqnarray}
\left(\begin{array}{ll}
r \\
z \\
\end{array}\right)_{\rm 4max} &=& \left(\begin{array}{ll}
0.71831446207852601 \\
0.98188047082901109 \\
\end{array}\right), \label{9.1} \\
\left(\begin{array}{ll}
r \\
z \\
\end{array}\right)_{\rm 4min} &=& \left(\begin{array}{ll}
0.92427265264269520 \\
0.97631849413594207 \\
\end{array}\right). \label{9.2}
\end{eqnarray}

3. On curves \# 6 to 8, $z(r)$ goes through a maximum, and then keeps decreasing
up to the edge of the model given by (\ref{4.3}). The coordinates of these
maxima are:
\begin{eqnarray}
\left(\begin{array}{ll}
r \\
z \\
\end{array}\right)_{\rm 5max} &=& \left(\begin{array}{ll}
0.65078589679780396 \\
0.95958866411532251 \\
\end{array}\right), \label{9.3} \\
\left(\begin{array}{ll}
r \\
z \\
\end{array}\right)_{\rm 6max} &=& \left(\begin{array}{ll}
0.64511331512911863 \\
0.95709853786786281 \\
\end{array}\right), \label{9.4} \\
\left(\begin{array}{ll}
r \\
z \\
\end{array}\right)_{\rm 7max} &=& \left(\begin{array}{ll}
0.63704541339084608 \\
0.95340148625320642 \\
\end{array}\right). \label{9.5}
\end{eqnarray}
These maxima \textit{do not} occur at the intersections of the corresponding
light cones with the SCS. At the SCS, $z(r)$ is already decreasing, which means
that light passing near an SCS acquires local blueshifts. The redshift profiles
remain smooth at those intersections and do not display any special behaviour
there.

\section{The extremum-redshift hypersurface}\label{maxred}

\setcounter{equation}{0}

Figure \ref{drawlatezall} shows that on the light cones that come near to the
SCS, the redshift begins to decrease on approaching the SCS, and, at the
intersection of the light cone with the SCS, $\dril z r < 0$. Thus, light
emitted close to the SCS displays local blueshifts, and this is an analogy to
the behaviour of light in a vicinity of a nonconstant BB
\cite{Szek1980,HeLa1984}, \cite{PlKr2006}.  The difference is that blueshifts
generated at the nonconstant BB are seen as infinite by all later observers,
while those generated at the SCS are finite, and, as the graphs in Fig.
\ref{drawlatezall} demonstrate, are swamped with excess by redshifts built up
later if the observer is far enough to the future. The central observer sees the
light from the SCS being redshifted.

The location of the hypersurface, at which blueshifts go over into redshifts
along radial rays, is observer-independent and can be calculated from
(\ref{2.7}). At that hypersurface, which we will call the extremum-redshift
hypersurface, we have $\dril z r = 0$, so $R,_{tr} = 0$. We find from
(\ref{3.11}):
\begin{eqnarray}\label{10.1}
R,_{tr} &=& \left[\sqrt{- k + {\cal F}} + \frac {r {\cal F},_r} {2 \sqrt{- k +
{\cal F}}}\right] \frac {\sinh \eta} {\cosh \eta - 1} \nonumber \\
&-& \frac 3 2\ \frac {r {\cal F},_r} {\sqrt{- k + {\cal F}}}\ \frac {\sinh \eta
- \eta} {(\cosh \eta - 1)^2}.
\end{eqnarray}
For numerical solving, the equation $R,_{tr} = 0$ can be written in a form
similar to (\ref{5.2}) -- (\ref{5.3}):
\begin{equation}\label{10.2}
\frac 1 r + \frac {{\cal F},_r} {2(-k + {\cal F})}\ [1 - 3 P(\eta)] = 0,
\end{equation}
where
\begin{equation}\label{10.3}
P(\eta) \df \frac {\sinh \eta - \eta} {\sinh \eta (\cosh \eta - 1)}.
\end{equation}
We have $P(0) = 1/3$, $\lim_{\eta \to \infty} P(\eta) = 0$ and $\dril P {\eta} <
0$ for all $\eta > 0$. Hence, $P(\eta)$ is monotonic in the full range of
$\eta$, and, if a solution of (\ref{10.2}) exists, then it is unique.

Having found $\eta$ for a given $r$, we then calculate $t(r)$ on the ERH from
(\ref{5.6}). The profile of the ERH is the lower ${\cal U}$-shaped curve in Fig.
\ref{shellcross}.

Using (\ref{5.2}) to eliminate ${\cal F},_r$, we find that at the SCS
\begin{equation}\label{10.4}
R,_{tr} = - \frac 3 2\ \frac {\sqrt{- k + {\cal F}} (\eta \cosh \eta + 2 \eta - 3
\sinh \eta)} {(\cosh \eta - 1)^2 (\tfrac 3 2 Q - 1)}.
\end{equation}
By (\ref{5.4}) -- (\ref{5.5}) we have $\tfrac 3 2 Q - 1 > 0$ for all $\eta
> 0$, and it is easy to verify that also $\eta \cosh \eta + 2 \eta - 3
\sinh \eta > 0$ for all $\eta > 0$. Therefore (\ref{10.4}) shows that $R,_{tr} <
0$ at the intersection of a radial light ray with the SCS, which, using
(\ref{2.7}), means $\dril z r < 0$. Hence, light emitted from the SCS is
blueshifted in a vicinity of the emission point. But the $R,_{tr}$ given by
(\ref{10.4}) is finite al all $\eta > 0$ (including $\eta \to \infty$, where
$R,_{tr} = 0$), so the blueshift is also finite.

The points on the light-cone profiles that correspond to the extrema in $z$
listed in (\ref{9.1}) -- (\ref{9.5}) should all lie on the ERH. They do, but
with a rather modest numerical precision. The $t$-coordinates of the points on
the ERH are smaller than the corresponding $t$-s on the light cones by $\approx
0.13$ NTU in the case of both extrema on curve \# 4 and of the maximum on curve
\#7, and by $\approx 0.14$ NTU in the other two cases.

We called the hypersurface given by (\ref{10.2}) -- (\ref{10.3}) an extremum
(rather than maximum) redshift hypersurface because, as seen from Fig.
\ref{drawlatezall}, on ray \# 4 that intersects this hypersurface twice, the
redshift profile has a maximum at one intersection point and a minimum at the
other. Generally, redshift profiles have maxima on the ``inside'' branch of the
ERH (the one closer to the center) and minima on the ``outside'' branch. On a
ray tangent to the ERH, the redshift profile has an inflection at the point of
tangency.

The appearance of local blueshifts is a signal that the ray (followed back in
time) is approaching the SCS. However, the blueshifts, being only local and
being overcompensated by redshifts along the later part of the ray, may not be
noticed by the future central observer, unless she is able to determine the
location of the light signal along the light cone independently of the value of
$z$, and with a sufficient precision. But even if she is, the precise moment of
crossing the SCS leaves no recognizable mark in the redshift profile -- $z(r)$
is smooth at the intersection of the ray with the SCS.

Note also that there are local blueshifts on curve \# 4 in Fig.
\ref{drawlatezall}, even though the corresponding ray never intersects the SCS
-- it just passes nearby. So, locating an SCS by observations is going to be a
difficult problem.

\section{The $D_A(z)$ and $D_L(z)$ relations on rays intersecting the
ERH}\label{dlz}

\setcounter{equation}{0}

\begin{figure}[h]
\includegraphics[scale=0.8]{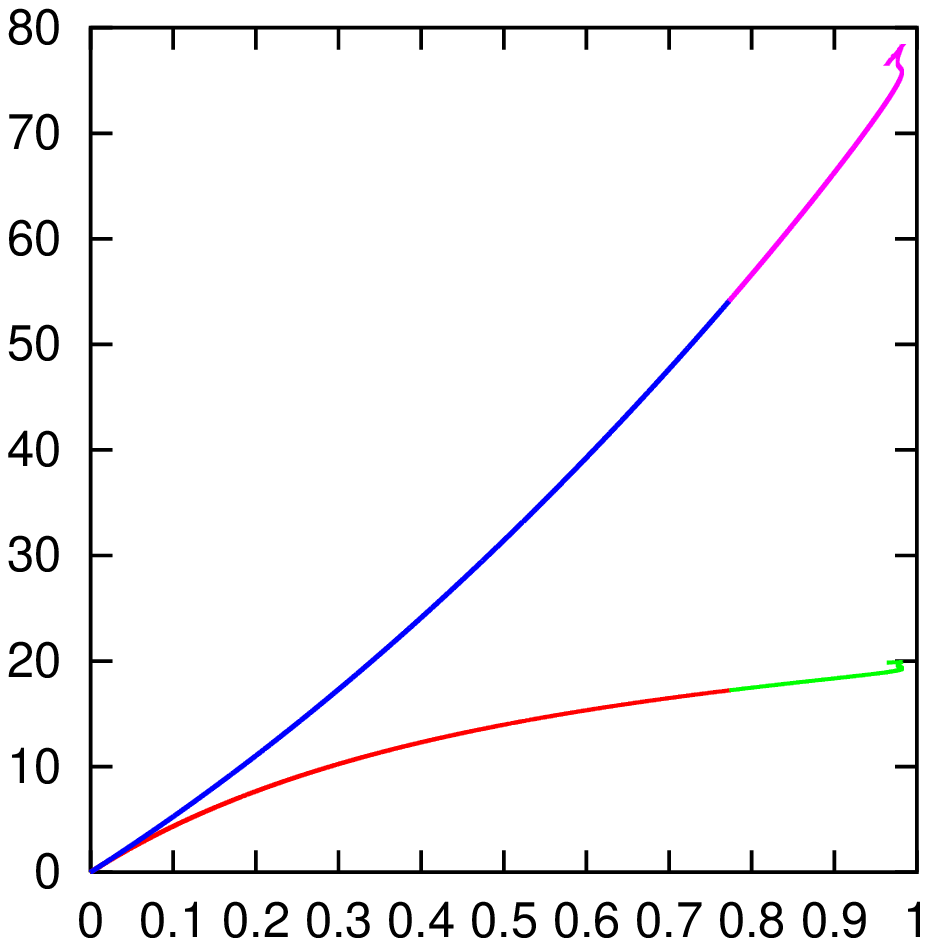}
${ }$ \\[-7.5cm]
\hspace{-3.5cm}
\includegraphics[scale=0.4]{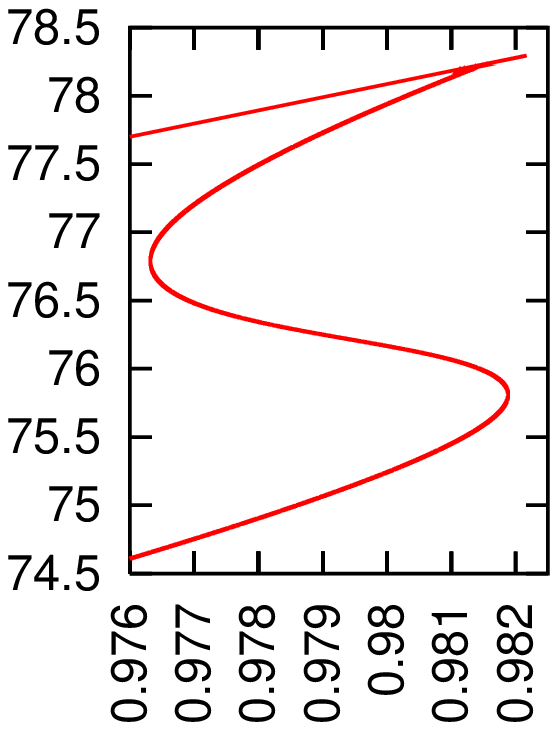}
${ }$ \\[-1.3cm]
\hspace{3mm}
\includegraphics[scale=0.4]{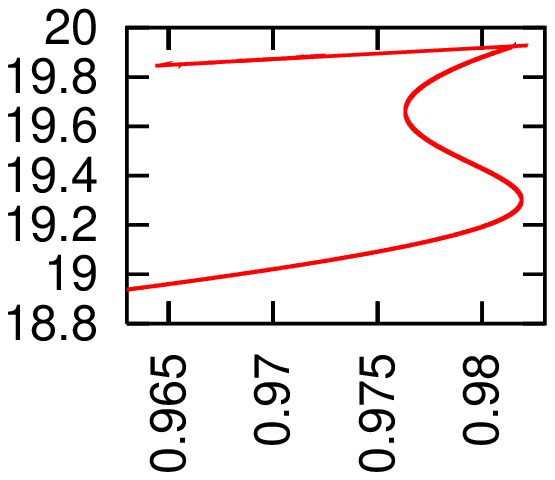}
\vspace{5cm} \caption{{\bf Main panel:} The relations $D_A(z)$ (lower curve) and
$D_L(z)$ (upper curve) along ray \# 4 from Fig. \ref{drawlatezall} (this is the
lowest ray in Fig. \ref{shellcross} that intersects the extremum-redshift
profile twice). {\bf Insets:} The upper ends of the curves from the main panel,
where the relations $D_L(z)$ (left inset) and $D_A(z)$ (right inset) become
triple-valued. The nearly straight segments at the upper edges of the graphs are
numerical fluctuations at the edge of the model.} \label{drawdisatt4}
\end{figure}

\begin{figure}[h]
\hspace{-3.2cm}
\includegraphics[scale=0.9]{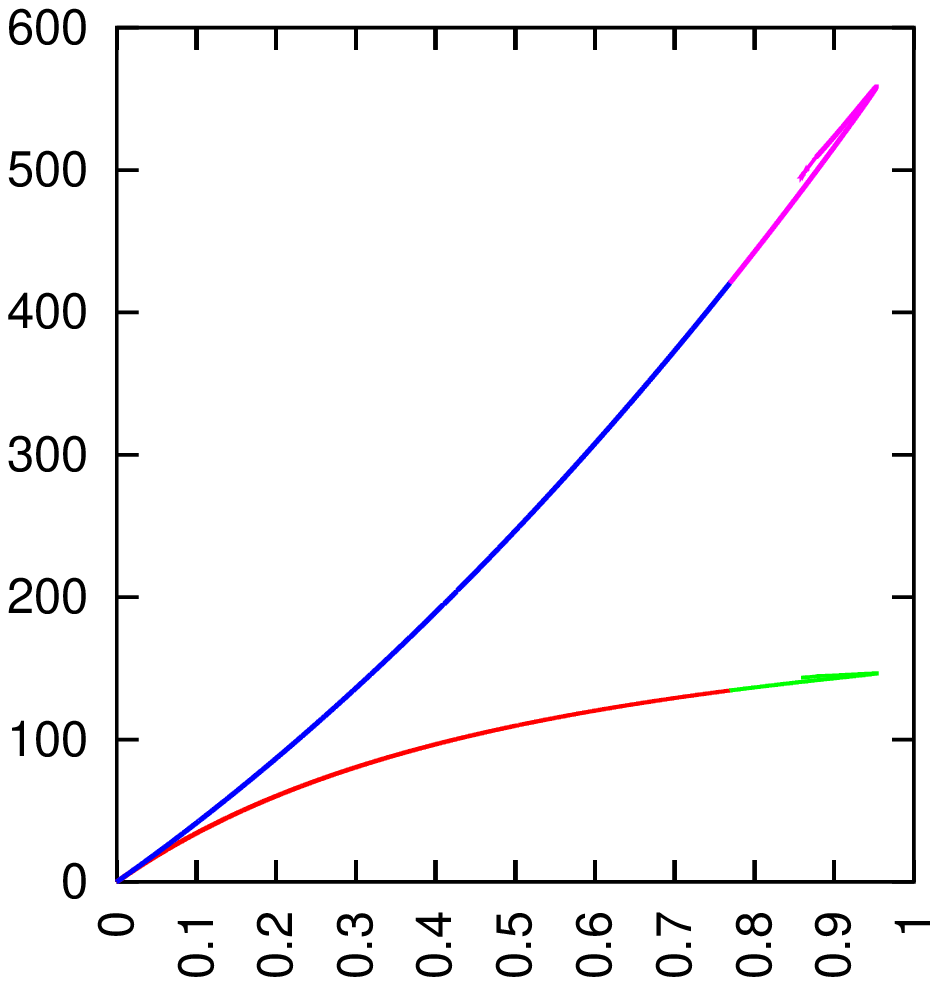}
${ }$ \\[-7cm]
\hspace{-3.3cm}
\includegraphics[scale=0.4]{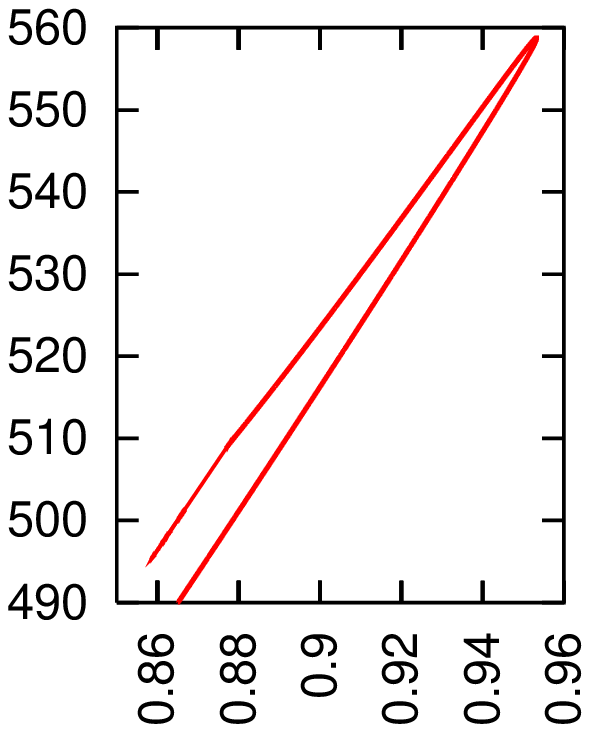}
${ }$ \\[-3cm]
\hspace{2cm}
\includegraphics[scale=0.4]{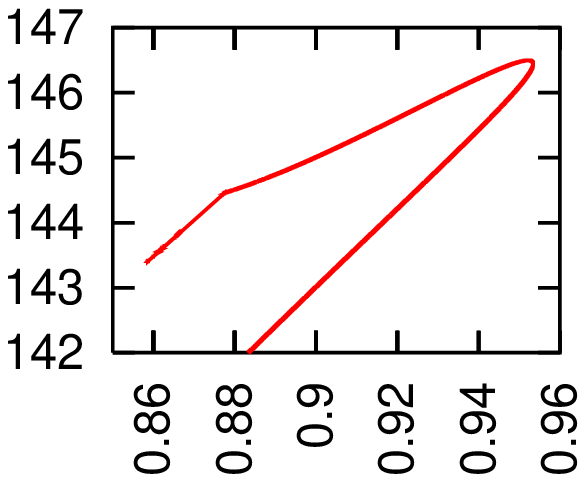}
\vspace{6.5cm}
\caption{The analogue of Fig. \ref{drawdisatt4} along ray \# 7 from Fig.
\ref{drawlatezall} (this is the highest ray in Fig. \ref{shellcross}). Now the
relations $D_L(z)$ and $D_A(z)$ become double-valued close to the shell
crossing.} \label{drawdisatt7}
\end{figure}

Quantities that are in principle observable, like the angular diameter distance
from the central observer $D_A$ (which is equal to the function $R(t_{\rm
ng}(r),r)$ appearing in (\ref{2.8})) or the luminosity distance from the central
observer $D_L$ (given by (\ref{2.8})), are usually expressed as functions of
redshift. The peculiar behaviour of redshift on rays passing near the SCS
($z(r)$ being non-monotonic) causes an even more peculiar behaviour of $D_A(z)$
and $D_L(z)$: where $z(r)$ is not monotonic, the relations $D_A(z)$ and $D_L(z)$
fail to be single-valued (and therefore cannot be called functions). This is
illustrated in Figs. \ref{drawdisatt4} and \ref{drawdisatt7} that show these two
relations along rays \# 4 and 7 from Fig. \ref{drawlatezall}. Along ray \# 4 in
the neighbourhood of the SCS, three values of $D_A$ and three values of $D_L$
correspond to a single value of $z$. Along ray \# 7 in the neighbourhood of the
SCS, the relations $D_A(z)$ and $D_L(z)$ become double-valued.

\section{The recombination epoch}\label{timing}

\setcounter{equation}{0}

It is interesting to compare some of the characteristics of the model discussed
here with those of the $\Lambda$CDM model. The metric of the $\Lambda$CDM model
is \cite{Kras2014}
\begin{equation}\label{12.1}
{\rm d} s^2 = {\rm d} t^2 - S^2(t) \left[{\rm d} r^2 + r^2 ({\rm d}\vartheta^2
+ \sin^2\vartheta \, {\rm d}\varphi^2)\right]
\end{equation}
with
\begin{equation}\label{12.2}
S(t) = \left(- \frac {6M_0} {\Lambda}\right)^{1/3} \sinh^{2/3} \left[\frac
{\sqrt {- 3 \Lambda}} 2 \left(t - t_{B\Lambda}\right)\right],
\end{equation}
where $M_0 = 1$ NLU, $t_B$ is given by (\ref{2.22}), and $\Lambda$ is calculated
from (\ref{2.10}) -- (\ref{2.13}), (\ref{2.18}) and (\ref{2.24}) to be
\begin{equation}\label{12.3}
- \Lambda = 3 \Omega_{\Lambda} {H_0}^2 = 91.849164\ ({\rm NLU})^{-2}.
\end{equation}
We assume
\begin{equation}\label{12.4}
t_{\rm rec} - t_B = 3.8 \times 10^5\ {\rm y} = 3.88 \times 10^{-6}\ {\rm NTU}
\end{equation}
for the time of last scattering \cite{swinxxxx}\footnote{In fact, the last
scattering was not an instant strictly localizable in time, but a process that
lasted some time (see e.g. Ref. \cite{Bole2006}). However, most astrophysical
papers do not take this into account.} in the $\Lambda$CDM model.

The mass density in all $p = 0$ RW models obeys \cite{PlKr2006}
\begin{equation}\label{12.5}
\kappa \rho = C_0 / S^3(t),
\end{equation}
where $C_0$ is a constant. Therefore, taking (\ref{6.4}) for the present mass
density, (\ref{2.21}) for the current value of $t - t_B$, and (\ref{12.4}) for
the value of $t - t_B$ at last scattering, we calculate the mass density
$\rho_{\rm ls}$ at last scattering to be
\begin{eqnarray}\label{12.6}
\kappa \rho_{\rm ls} &=& \kappa \rho_0\ \frac {S^3(T)} {S^3(t_{\rm rec} - t_B)}
\\
&=& 88089589221.4818 \approx 88 \times 10^9\ ({\rm NLU})^{-2}. \nonumber
\end{eqnarray}

We will now assume that the last scattering in the L--T model occurs at the same
density; the last scattering time will thus depend on $r$. This is an
approximate method of determining $(t_{\rm rec} - t_B)$; for a more precise
method see Refs. \cite{YNSa2010} and \cite{YNSa2010a}.

A density approximately equal to (\ref{12.6}),\footnote{The values that follow
were read off from the numerical tables used to draw the figures in this paper.}
namely
\begin{equation}\label{12.7}
\kappa \rho_{\rm LTLS} = 88017457848.852432,
\end{equation}
is attained along the PCPO at the redshift
\begin{equation}\label{12.8}
z_{\rm LTLS} = 1054.891484271654,
\end{equation}
not much different from (\ref{7.3}). This happens at
\begin{equation}\label{12.9}
r_{\rm LTLS} = 1.010092188377007,
\end{equation}
and this corresponds to the time along the PCPO
\begin{equation}\label{12.10}
t_{\rm LTLS} = -0.1329399592464457,
\end{equation}
which is later than the BB by
\begin{equation}\label{12.11}
\tau = 3.3614380286 \times 10^{-6}\ {\rm NTU} \approx 3.2942 \times 10^{5}\
{\rm y}.
\end{equation}

\begin{figure}[h]
\hspace{-5mm}
\includegraphics[scale=0.5]{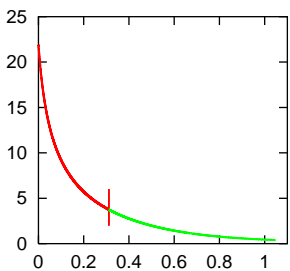}
\caption{Graph of the function $(- k_F(r)) \equiv |- k + {\cal F}|$. The
vertical stroke is at $r = r_{\rm AH}$. } \label{kequiv}
\end{figure}

The difference between (\ref{12.11}) and (\ref{12.4}) has a simple intuitive
explanation (see Fig. \ref{kequiv}, copied from Ref. \cite{Kras2014}). In any
L--T model, every constant-$r$ shell evolves by eq. (\ref{2.2}), which is the
same as the Friedmann equation, except that here $E(r)$ is different for every
shell, and so is $M(r)$. The Friedmann curvature index $k_F$ is related to $E$
by $2E = - k_F r^2$, so, by (\ref{3.2}), we have
\begin{equation}\label{12.12}
k_F(r) = k - {\cal F}(r),
\end{equation}
i.e. it is different for every shell. As Fig. \ref{kequiv} shows, $|k_F|$ is
largest at $r = 0$ and monotonically decreasing, so the matter shells closer to
$r = 0$ expand according to a ``faster'' Friedmann equation than the more
distant ones. Also, they expand faster than the corresponding $\Lambda$CDM
shells, as can be seen by comparing (\ref{2.21}) with (\ref{2.22}) and
(\ref{6.1}) with (\ref{6.3}): the central region of the L--T model produced a
lower density than $\Lambda$CDM in a shorter time. Consequently, the density
required for recombination must have been achieved in a shorter time, too.

\section{Final summary}\label{conclu}

\setcounter{equation}{0}

The L--T model considered here duplicates the luminosity distance vs. redshift
relation $D_L(z)$ of the $\Lambda$CDM model using the energy function $E(r)$
alone. Its existence was proved first in Ref. \cite{INNa2002}, and then, by a
more precise method, in Ref. \cite{YKNa2008}. It was explicitly (numerically)
constructed in Refs. \cite{YKNa2008} and (in different coordinates)
\cite{Kras2014}. It turned out that this model necessarily contains a region
where shell crossings occur \cite{Kras2014}. This region is far enough from the
central observer to cause no problems with the interpretation of observations of
the type Ia supernovae -- its inner boundary intersects the past light cone of
the present observer at $z \approx 6.9$. Therefore, it can be removed from the
spacetime by matching the L--T model to a Friedmann background, without harming
the usability of the model for explaining the $D_L(z)$ function. This much was
proven in Ref. \cite{Kras2014}.

Here, the consequences of existence of the shell crossings were investigated.
They are interesting from the point of view of geometry. They might become
meaningful also for cosmology if a qualitatively similar model with different
numerical parameters emerges from some other research. In such a model, the SCS
might appear early enough to be visible for the present observer.

The model extends from the center of symmetry up to a finite distance (Sec.
\ref{lightcone}). The edge of the model is formed by the world lines of those
cosmic matter particles that were ejected from the Big Bang at its contact with
the past light cone of the present central observer. The model can be extended
by matching it to either another L--T model or to a Friedmann model, but the
extensions are arbitrary -- they are not constrained by the $D_L(z)$ relation
that defined our model, and were not considered in this paper.

The earliest point of the shell-crossing set is at $t = t_{\rm min} \approx
74.8$ NTU $\approx 562.4 T$ to the future from now, where $T = 13.819 \times
10^9\ {\rm y} = 0.141\ {\rm NTU}$ is the present age of the Universe (Sec.
\ref{locateSCS}). The signal sent from that point would reach the central
observer at $t_F \approx 149.965\ {\rm NTU} \approx 1063.58 T$ to the future
from now (Sec. \ref{SCSinflu}).

Mass density distributions along a few hypersurfaces of constant $t$, and along
the past light cone of the present central observer were numerically calculated
(Secs. \ref{densities} and \ref{densoncone}). As expected, the density goes to
infinity wherever such a hypersurface touches or intersects the SCS.

Characteristic examples of light rays that intersect the SCS were calculated
(Sec. \ref{SCSinflu}). As expected, in the comoving coordinates they are
horizontal at the intersection points. Then, redshift profiles along several
light rays were calculated, including those mentioned above (Sec.
\ref{redprof}). It turned out that on rays passing near the SCS, redshift
acquires a maximum \textit{before} the ray crosses the SCS, and at the
intersection with the SCS the function $z(r)$ is already decreasing. It is
surprising that the intersection with the SCS leaves no recognizable trace in
the redshift profile: it is smooth there, and has no extremum. Thus, an observer
placed down a light cone from the SCS would not notice any sign of the mass
density being infinite there.

The extremum-redshift hypersurface, on which the redshift profiles acquire
maxima and minima, was determined in Sec. \ref{maxred}. Rays that cross the ERH
display local blueshifts on the other side of it. In particular, they are
blueshifted when emitted at the SCS. However, the blueshifts are finite, and
become overcompensated by redshifts before the ray reaches the central observer.

The relations $D_A(z)$ and $D_L(z)$ along two rays (one passing near, the other
crossing the SCS) were displayed in Sec. \ref{dlz} ($D_A$ is the angular
diameter distance and $D_L$ is the luminosity distance from the central
observer). In consequence of the blueshifts generated in a vicinity of the SCS,
these relations become double- or triple-valued near the SCS.

Finally, the end-instant of the recombination epoch along the radial ray
reaching the central observer at present was calculated for the model considered
here. It occurs at $t \approx 3.29 \times 10^5$ y after the Big Bang, vs. $3.8
\times 10^5$ y in the $\Lambda$CDM model. The difference finds a simple
explanation in the profile of the $E(r)$ function; details are given in Sec.
\ref{timing}.

It is hoped that the investigation presented here will be useful if a model with
a shell crossing within the view of the present observer emerges from some
future research.

\appendix*

\section{The proof of (\ref{5.5})} \label{prove48}

\setcounter{equation}{0}

The function $Q(\eta)$ of (\ref{5.5}) can be written as

\begin{equation}\label{a.1}
Q(x) = \frac {\cosh^2 x} {\sinh^2 x} - \frac {x \cosh x} {\sinh^3 x}, \quad {\rm
where} \quad x \df \eta/2.
\end{equation}
Then
\begin{equation}\label{a.2}
\dr Q x = \frac {L(x)} {\sinh^4 x},
\end{equation}
where
\begin{equation}\label{a.3}
L(x) \df - 3 \sinh x \cosh x + 2 x \sinh^2 x + 3x.
\end{equation}
The $L(x)$ has the following properties
\begin{eqnarray}
&& L(0) = 0, \label{a.4} \\
&& \dril L x = - 3 \cosh^2 x - \sinh^2 x + 4 x \sinh x \cosh x + 3, \nonumber
\\
\label{a.5} \\
&& (\dril L x)(0) = 0, \label{a.6} \\
&& \dril {^2 L} {x^2} = - 4 \sinh x \cosh x + 4 x (\sinh^2 x + \cosh^2 x),
\nonumber \\
\label{a.7} \\
&& (\dril {^2 L} {x^2})(0) = 0, \label{a.8} \\
&& \dril {^3 L} {x^3} = 16 \sinh x \cosh x > 0 \quad {\rm for\ all} \quad x > 0.
\label{a.9}
\end{eqnarray}
Equations (\ref{a.4}) -- (\ref{a.9}) considered in reverse order imply that
$L(x) > 0$ for all $x > 0$, so $\dril Q {\eta} = 2 \dril Q x > 0$ for all $\eta
> 0$. $\square$

\bigskip

{\bf Acknowledgements} I am grateful to Jim Zibin for an insightful remark, and
to the referee for comments that inspired an extension of the first version of
this paper.

\end{document}